\documentclass[prc,aps,a4paper,groupedaddress,superscriptaddress,nofootinbib,showpacs
,preprintnumbers,onecolumn]{revtex4}
\usepackage{graphicx}
\usepackage{amsfonts,ulem}
\usepackage{amssymb}
\usepackage{natbib}
\usepackage{color}
\usepackage{dcolumn}
\usepackage{bm}
\usepackage{subeqnarray}
\newcommand{\bwt}{\begin{widetext}}
\newcommand{\ewt}{\end{widetext}}
\newcommand{\beq}{\begin{equation}}
\newcommand{\eeq}{\end{equation}}
\newcommand{\bea}{\begin{eqnarray}}
\newcommand{\eea}{\end{eqnarray}}
\newcommand{\upa}{\uparrow}
\newcommand{\dwa}{\downarrow}
\begin{document}
\title{Magnetic susceptibility and magnetization properties of asymmetric nuclear matter under a strong magnetic field}
\author{A.~Rabhi}
\email{rabhi@teor.fis.uc.pt}
\affiliation{Centro de F\' {\i}sica Computacional, Department of Physics, University of Coimbra, 3004-516 Coimbra, Portugal} 
\affiliation{University of Tunis El-Manar, Unit\'e de Recherche de Physique Nucl\'eaire et des Hautes \'Energies, Facult\'e des Sciences de Tunis, 2092 Tunis, Tunisia}
\author{M.~A.~P\'erez-Garc\'{\i}a}
\email{mperezga@usal.es}
\affiliation{Departamento de F\'{\i}sica Fundamental and IUFFyM, Universidad de Salamanca, Salamanca, Spain.}
\author{C.~Provid\^encia}
\email{cp@teor.fis.uc.pt}
\affiliation{Centro de F\' {\i}sica Computacional, Department of Physics, University of Coimbra, 3004-516 Coimbra, Portugal} 
\author{I.~Vida\~na}
\email{ividana@fis.uc.pt}
\affiliation{Centro de F\' {\i}sica Computacional, Department of Physics, University of Coimbra, 3004-516 Coimbra, Portugal} 

\date{\today}
\begin{abstract} We study the effect of a strong magnetic field on  the proton and neutron 
  spin polarization and magnetic susceptibility
  of asymmetric nuclear matter within a relativistic mean-field
  approach. It is shown that  magnetic fields 
  $B\sim 10^{16}$ - $10^{17}$ G have already noticeable effects on the range
  of densities of interest for the study of the crust of a neutron
  star. Although the proton susceptibility is larger for weaker
  fields, the neutron susceptibility becomes of the same order or even 
  larger for small proton fractions and subsaturation densities for
  $B>10^{16}$ G. { We expect that  neutron superfluidity in the crust will  be
affected by 
the presence of magnetic fields.}
\end{abstract}
\maketitle

\section{Introduction}

A well grounded understanding of the properties of isospin asymmetric nuclear systems, such as nuclei far from the stability valley or astronomical objects like neutron stars, is crucial for the advancement of both nuclear physics and astrophysics \cite{epja,baran, baoan,steiner}. A major experimental and theoretical effort with significant progress has been carried out over the last years (see {\it e.g.,} Ref.\ \cite{epja} and references therein) in the study of the properties of isospin-asymmetric nuclear systems. Laboratory experiments, such as those recently performed or being planned in existing or next-generation radioactive ion beam facilities, are providing crucial information on the isospin dependence of the nuclear force from a large number of heavy ion collision and
nuclear structure observables.  Complementary information can be extracted from the observation of neutron stars, which open a window into both the bulk and microscopic properties of nuclear matter at extreme isospin asymmetries. 

Neutron stars, most of which are detected as pulsars, have strong surface magnetic fields which can reach values of the order of $10^{14}-10^{15}$ G in the case of the so-called magnetars\footnote{for a review on magnetars see {\it e.g.,} Ref.\ \cite{harding06}} that may grow by several orders of magnitude in dense interior of the star. Despite the great theoretical effort of the last forty years, there is still no general consensus regarding the mechanism to generate such strong magnetic fields in a neutron star. The field could be a fossil remnant from that of the progenitor star \cite{tatsumi06}, or alternatively, it could be generated after the formation of the neutron star by some long-lived electric currents flowing in the highly conductive neutron star material \cite{thompson93}. From the nuclear physics point of view, however, one of the most interesting and stimulating mechanisms that has been suggested is the possible existence of a phase transition to a ferromagnetic state at densities corresponding to the theoretically stable neutron stars and, therefore, of a ferromagnetic core in the liquid interior of such compact objects. Such a possibility has long been considered by many authors within different theoretical approaches (see {\it e.g.,}
\cite{BROWNELL,RICE,CLARK,CLARK2,SILVERSTEIN,OST,PEAR,PANDA,BACK,HAENSEL,JACK,KUT,MARCOS,
VIDAURRE,RIOS,GOGNY,FANTONI,VIDANA,VIDANAb,VIDANAc,SAMMARRUCA1,SAMMARRUCA2,BIGDELI,ANG09a,ANG09b}),
but results were contradictory. Therefore, a complete understanding of
the magnetic properties of neutron stars and, more generally, of that
of isospin asymmetric nuclear matter, requires the study of nuclear
matter under the influence of magnetic fields. 

{ An estimation of the maximum magnetic field intensity supported
  by a star before magnetic field stresses give rise to the formation
  of a black hole may be obtained equating the magnetic field energy
  of an uniform field in a sphere with the star radius $R$
  to the gravitational binding energy. Using the Schwarzschild
  criterion that $R> 2MG/c^2$, the magnetic field should satisfy $B\le
  8\times 10^{18}\left( 1.4 M_\odot/M\right)$G \cite{cardall2001}.
  This estimation is just
slightly larger than the maximum fields obtained in the
 framework of a
 relativistic magneto-hydrostatic formalism \cite{cardall2001}, which were of
the order of $\sim 5\times 10^{18}$ G. In \cite{bpl02}
 using the same approach and an hyperonic
EOS the authors have obtained stable configurations for $B\le 3 \times
10^{18}$ G.  However, they  did not exclude larger fields, and, in
particular, they suggest that a non constant current function with an appreciable
gradient or a disordered field with $\langle B^2\rangle> \langle \vec
B\rangle^2$ could possibly give rise to larger fields in still stable stars. Taking these numbers as indicative we
will consider fields $B\le
10^{19}$ G in the present work.}

Until presently fields no higher than $\sim 10^{15}$ G have been
measured at the surface of magnetars, see  \cite{mcgill} \footnote{http://www.physics.mcgill.ca/\~{}pulsar/magnetar/main.html}.  From the
 observation of the spin-down of pulsars via electromagnetic radiation the  surface poloidal magnetic field can be estimated. However, there
 is some evidence that the internal magnetic field of several external
 low-field magnetars could be  higher. An example is the SGR 0418+5729 whose dipole field is less
 than a few times $10^{12}$ G: an internal toroidal field at large as $\sim 10^{16}$G could have existed inside its interior in order to explain some of the presently observed properties \cite{turolla2011}. Using a pure toroidal
 magnetic field equilibrium  models of relativistic stars have been
 calculated for both non-rotating and rotating stars. These
 toroidal fields vanish at the symmetry axis, have a maximum value,
 that can be larger than $10^{18}$ G depending on the EOS, on the 
 equatorial plane deep inside the star and decreases toward the surface where it 
 vanishes \cite{rezzolla2012}.
   Let us also point out that the internal poloidal fields calculated using a
 general relativistic magneto-hydrostatic formalism are larger than the 
 surface ones \cite{oertel2014}. It has also been suggested that magnetic fields 
 may be held in the core for periods much longer than the ohmic
 diffusion time due to interactions between the magnetic flux tubes
 and the  vortex tubes expected to be present in a superconducting, superfluid, 
 rotating neutron star \cite{ruderman1998}. The detailed underlying mechanism for this is, however, far from being completely understood.

Particularly interesting is the study of the magnetization of matter due to the presence of a magnetic field. Whereas the magnetization of symmetric nuclear matter and pure neutron matter has been studied by several authors \cite{kha2002,APG08,APG11,aguirre2011,Aguirre13,abv2014}, the magnetization of $\beta$-stable neutron star matter has, however, received less attention in the literature. In Ref.~\cite{bh82}, for instance, the magnetization of $\beta$-stable matter was extensively studied for a single component electron gas and for the crust matter of neutron star. This study was latter generalized by Broderick {\it et al.} \cite{bpl00} by including also the contribution of neutrons and protons. Recently, Dong {\it et al.,} \cite{dzg2013} have studied the effect of density dependence of the nuclear symmetry energy on the magnetization of $\beta$-stable matter. These authors concluded that the magnetic susceptibility of charged particles (protons, electrons and muons) can be larger than that of the neutron, and that the anomalous magnetic moment of the protons enhances their magnetic susceptibility to the point that it can be one of the main contributions and, therefore, should not be neglected. They also show that the proton magnetic susceptibility is sensitive to the density dependence of the nuclear symmetry energy, namely to the isospin content of the nuclear force.  

In this work we study the magnetization of spin polarized isospin
asymmetric nuclear matter at zero temperature by using a relativistic
mean field  approach. The scope of this work is threefold: (i) to
determine under which conditions of density, isospin asymmetry matter
and magnetic fields matter is totally polarized, (ii) to compare under
such conditions the proton and neutron magnetic susceptibilities, and
finally (iii) to determine which is in each case the most
energetically favorable spin configuration. The density dependence of
the energy of the system and its pressure, as well as its
compressibility is analyzed for different proton fractions and
magnetic fields.  { We will not consider
  $\beta$-equilibrium matter in most of the results shown, but will consider a wide range of nuclear matter asymmetries 
of interest for stellar matter, in particular, to the study of the inner crust,
where a pasta phase is expected \cite{lima2013}, to the study of matter with trapped neutrinos
where large proton fractions are expected, which may
be as large as 0.4 in the presence of a magnetic field
\cite{rabhi2010}, and to the study of neutrino free  matter in
$\beta$-equilibrium where the proton fraction will increase above 0.1 at subsaturation
densities for a strong enough magnetic field \cite{rpp08}.  
Therefore, besides symmetric
nuclear matter and neutron matter, we will choose two representative proton
  fractions  namely $Y_p=0.1$ for cold $\beta$-equilibrium matter, and
  $Y_p=0.3$ for warm protoneutron star matter with a fraction of
  0.4 trapped leptons. For reference, we will also present the proton
  fraction of $\beta$-equilibrium matter for the magnetic field
  intensities considered in the present work, as well as the
  proton and neutron polarization and magnetization of $\beta$-equilibrium matter for some of the cases discussed. Let us point out that, even
though a protoneutron star should be described at finite temperature,
which has as immediate consequence the wash out of Landau levels, the
main features defined by a large isospin symmetry due to  neutrino
trapping may be understood at zero temperature.  
}
In addition, the role of the proton
anomalous magnetic moment is investigated in detail.

The paper is organized in the following way. A short review of the formalism is presented in Sec.\ \ref{section1}. In Sec.\ ~\ref{section2} we present explicit expressions for the magnetization of each nucleonic species, as well as for their differential susceptibilities. The results are shown and discussed in Sec.\ \ref{section3}.  Finally, a short summary and our main conclusions are given in Sec.\ \ref{section4}. 

 
\section{The formalism}
\label{section1}
To describe nuclear matter in a external uniform magnetic field $B$ along the $z$-axis, we employ a relativistic mean field (RMF) approach, in which the nucleons interact via the exchange of $\sigma$, $\omega$ and $\rho$ mesons. The total interacting Lagrangian density of the non-linear Walecka model (NLWM) has the form
\beq
{\cal L}= \sum_{N=n, p}{\cal L}_{N} + {\cal L}_{m}.
\label{lan}
\eeq
The nucleon ($N$=$n$, $p$) Lagrangian density, including meson-nucleon interacting terms, and the meson
($\sigma$, $\omega$ and  $\rho$) Lagrangian density are, respectively, given by 
\bwt
\bea
{\cal L}_{N}&=&\bar{\Psi}_{N}\left(i\gamma_{\mu}\partial^{\mu}-q_{N}\gamma_{\mu}A^{\mu}- 
m+g_{\sigma}\sigma
-g_{\omega}\gamma_{\mu}\omega^{\mu}-\frac{1}{2}g_{\rho}\boldsymbol{\tau}.\gamma_{\mu}\boldsymbol{\rho}^{\mu}
-\frac{1}{2}\mu_{N}\kappa_{N}\sigma_{\mu \nu} F^{\mu \nu}\right )\Psi_{N} 
\label{lagran}
\eea
\ewt
and
\bwt
\bea
{\cal L}_{m}&=&\frac{1}{2}\partial_{\mu}\sigma \partial^{\mu}\sigma
-\frac{1}{2}m^{2}_{\sigma}\sigma^{2}-\frac{1}{3!}c_{\sigma} \sigma^{3} -\frac{1}{4!}\lambda \sigma^{4} 
+\frac{1}{2}m^{2}_{\omega}\omega_{\mu}\omega^{\mu}+\frac{1}{4!}\xi g^4_\omega(\omega_{\mu}\omega^{\mu} )^2
-\frac{1}{4}\Omega^{\mu \nu} \Omega_{\mu \nu}  \cr
&-&\frac{1}{4} F^{\mu \nu}F_{\mu \nu}
+\frac{1}{2}m^{2}_{\rho}\boldsymbol{\rho_{\mu}}.\boldsymbol{\rho^{\mu}}-\frac{1}{4}  P^{\mu \nu}P_{\mu \nu} + \Lambda_{\omega} g_\rho^2\boldsymbol{\rho}_{\mu}.\boldsymbol{\rho}^{\mu} g_{\omega}^2\omega_{\mu}\omega^{\mu}\ .
\label{lagran2}
\eea
\ewt
In the above expressions, $\Psi_{N}$ are the nucleon Dirac fields, the nucleon mass is denoted by $m$, $\boldsymbol{\tau}$ are the isospin Pauli matrices, and
$\mu_N$ is the nuclear magneton.  The nucleon anomalous magnetic (AMM) moments are introduced via the coupling of the baryons to the electromagnetic field tensor with $\sigma_{\mu \nu}=\frac{i}{2}\left[\gamma_{\mu}, \gamma_{\nu}\right] $ and strength
$\kappa_{N}$, with $\kappa_{n}=-1.91315$ for the neutron and $\kappa_{p}=1.79285$ for the proton, respectively. 
The mesonic and electromagnetic field strength tensors are given by their usual expressions: $\Omega_{\mu \nu}=\partial_{\mu}\omega_{\nu}-\partial_{\nu}\omega_{\mu}$, $P_{\mu 
\nu}=\partial_{\mu}\boldsymbol{\rho}_{\nu}-\partial_{\nu}\boldsymbol{\rho}_{\mu}-g_\rho(\boldsymbol{\rho}_\mu \times \boldsymbol{\rho}_\nu)$, and  $F_{\mu
\nu}=\partial_{\mu}A_{\nu}-\partial_{\nu}A_{\mu}$. The photon field  $A^\mu$ is taken as $(0,0,Bx,0)$ in such a way that the external magnetic field $\vec B$ is aligned with the $z$-axis.  The electromagnetic field is assumed to be externally generated (and thus
has no associated field equation), and only frozen-field configurations will be considered. The
nucleon-meson couplings are denoted by $g$ and the electromagnetic
couplings by $q$. The parameters of the model are the nucleon mass $m_N$,
the masses of mesons $m_\sigma$, $m_\omega$, and $m_\rho$, and the
nucleon-meson couplings. The self-interaction term with coupling
constants $c_{\sigma}$ and $\lambda$ for the $\sigma$ meson are
introduced. The RMF parametrization employed in this work is the
FSUGold \cite{fsugold} where two more parameters $\xi$ and $\Lambda_\omega$ have been
introduced: $\xi$ to describe the $\omega-$meson self-interactions,
which softens the equation of state  at high density, and
$\Lambda_\omega$, a nonlinear mixed isoscalar-isovector term which
modifies the dependence of the symmetry energy.  The FSUGold model has been chosen
because it is frequently
applied in the description of nuclear matter and stellar hadronic
matter {\cite{fsu2}}. Although, FSUGold is too soft at large densities and it is not capable of
describing a 2 $M_\odot$ neutron star, we expect it will describe well nuclear
matter below $3\rho_0$, the range of densities we will analyze.

From now we take the standard mean-field theory approach and display only some of the equations needed for this study. A complete set of equations and description of the method can 
be found in the literature (\textit{e.g.}, Ref.~\cite{cbp97,bpl00,rpp08}). For
the description of the system, we need the energy density of nuclear
matter, the pressure and the baryonic density.
The energy density of nuclear matter can be expressed as
\beq
\varepsilon=\varepsilon_{n}+\varepsilon_{p}+\frac{1}{2}m^{2}_{\sigma}\sigma^2
+\frac{1}{3!}c_\sigma g_\sigma^{3}\sigma^{3}+\frac{1}{4!}\lambda g^{4}_{\sigma}\sigma^{4}
+\frac{1}{2}m^{2}_{\omega}\omega^{2}_{0}+\frac{1}{8}\xi g^{4}_{\omega}\omega^{4}_{0}
+\frac{1}{2}m^{2}_{\rho}\rho^{2}_{0}
+3\Lambda_{\omega}g^{2}_{\rho}\rho_{0}^{2}
g^{2}_{\omega}\omega^{2}_{0},
\label{eq:ed}
\eeq
and the pressure of the system is obtained from the thermodynamical relation
\beq
P_{m}=\sum_{i=n,p}\mu_{i}\rho_{i}-\varepsilon\ .
\label{press}
\eeq
The proton and neutron chemical potentials read
\bea
\mu_p &=& E^p_F+g_\omega\omega^0+\frac{1}{2}g_\rho\rho^0 \\
\mu_n &=& E^n_F+g_\omega\omega^0-\frac{1}{2}g_\rho\rho^0,
\eea
where $E^p_F$ and $E^n_F$ are the proton and neutron Fermi energies related to their corresponding Fermi momenta $k^p_{F, \nu, s}$ and  $k^{n}_{F, s}$ through
\bea
\Big(k^{p}_{F,\nu, s}\Big)^2&=& \Big(E^{p }_{F}\Big)^2-\left[\sqrt{m^{*2}+2\nu q_pB}-s\mu_N\kappa_pB \right]^{2} \\
\Big(k^{n}_{F, s}\Big)^2 &=& \Big(E^{n}_{F}\Big)^2-\bar{m}^{2}_{ns} \,  
\eea
where $\nu=n+\frac{1}{2}-sgn(q)\frac{s}{2}=0, 1, 2, \ldots$ enumerates
the Landau levels of the fermions with electric charge $q$, the
quantum number $s$ is $+1$ for spin up ($\upa$) and $-1$ for spin down ($\dwa$)
particles, and for the neutrons we have introduced 
\beq
\bar{m}_{ns}=m^{*}-s\mu_N\kappa_n B, 
\eeq
with $m^*$ the nucleon effective mass given by
\beq
m^{*}=m-g_\sigma \sigma \ .
\eeq
The proton and neutron densities are given by
\bea
\rho_{p}&=&\frac{q_{p}B}{2\pi^{2}}\sum_{\nu, s}k^{p}_{F,\nu,s} \\
\rho_{n}&=&\frac{1}{2\pi^{2}}\sum_{s}\left[ \frac{1}{3}\left(k^{n}_{F, s}\right) ^{3}-\frac{1}
{2}s\mu_{N}\kappa_{n}B\left(\bar{m}_{ns}k^{n}_{F,s}+\Big(E^{n}_{F}\Big)^2\left(\arcsin\left( \frac{\bar{m}_{ns}}
{E^{n}_{F}}\right) -\frac{\pi}{2} \right)  \right) \right] \, 
\eea
where the summation over the index $\nu$ in the expression for the proton density starts from 0 (1) for spin-up (spin-down) protons and runs up to the largest integer for which the 
square of the Fermi momentum of the proton is still positive. This maximum value of $\nu$ is defined by the ratio
\beq
\nu_{max}=\left[\frac{\left(E^{p}_{F}+s\mu_{N}\kappa_{p}B\right)^2-m^{*2}}{2|q_{p}|B}\right].
\eeq

Finally, the proton and neutron energy densities $\varepsilon_p$ and $\varepsilon_n$
that enter the total energy density (\ref{eq:ed}) are given by
\bea
\varepsilon_{p}&=&\frac{q_{p}B}{4\pi^ {2}}\sum_{\nu, s}\left[k^{p}_{F,\nu,s}E^{p}_{F}
+\left(\sqrt{m^{* 2}+2\nu q_{p}B}-s\mu_{N}\kappa_{p}B\right) ^{2} 
\ln\left|\frac{k^{p}_{F,\nu,s}+E^{p}_{F}}{\sqrt{m^{* 2}+2\nu q_{p}B}-s\mu_{N}\kappa_{p}B} \right|\right]  \\
\varepsilon_{n}&=&\frac{1}{4\pi^ {2}}\sum_{s}\bigg[\frac{1}{2}k^{n}_{F, s}\Big(E^{n}_{F}\Big)^3-\frac{2}
{3}s\mu_{N}\kappa_{n} B \Big(E^{n}_{F}\Big)^3\left(\arcsin\left(\frac{\bar{m}_{ns}}{E^{n}_{F}} \right)-\frac{\pi}
{2}\right)-\left(\frac{1}{3}s\mu_{N}\kappa_{n} B +\frac{1}{4}\bar{m}_{ns}\right) \cr
&&\left(\bar{m}_{ns}k^{n}_{F, s}E^{n}_{F}+\bar{m}^{3}_{ns}\ln\left|\frac{k^{n}_{F,s}+E^{n}_{F}}{\bar{m}_{ns}} 
\right|\right) \bigg],
\eea

\section{Magnetic susceptibility}
\label{section2}
The magnetization of nuclear matter defined as the derivative of the energy
density with respect to B, at constant baryonic density $\rho$ and fixed
proton fraction $Y_p$, can be written as
\beq
{\cal M}=-\left.\frac{\partial \varepsilon}{\partial B}\right|_{\rho, Y_p}=\sum_{i=p,n}
\left.-\frac{\partial \varepsilon_i}{\partial B}\right|_{\rho, Y_p} =\sum_{i=p,n}{\cal M}_i.
\eeq
Note that, since the density and the proton fraction is considered fixed, there is no contribution  from the meson fields to the magnetization in this case. 
The proton magnetization, ${\cal M}_p$, is given by \cite{bpl00, HHRS10, dzg2013}
\beq
{\cal M}_p=-\frac{\varepsilon_p}{B}+E^p_F\frac{\rho_p}{B}-
\frac{q_p B}{2\pi^2}\sum_{\nu,s} \bar m_{p\nu s}\ln\left|\frac{ E^p_F+k^p_{F,\nu, s}}{\bar
m_{p\nu s}}\right|\left(\frac{q_p\nu}{\tilde m_{p\nu s}}-s\mu_N\kappa_p\right),
\eeq
where $\bar m_{p\nu s}$ is defined as
\beq
\bar m_{p\nu s}=\tilde m_{p\nu}-s\mu_N\kappa_p B
\eeq
with 
\beq
\tilde m_{p\nu}=\sqrt{m^{*2}+ 2\,q_p\,\nu B} \ .
\eeq
The magnetization of the neutrons ${\cal M}_n$ reads
\bea
{\cal M}_n &=&\frac{1}{2\pi^2}\sum_{s} \left(s\mu_N\kappa_n\right) \left\lbrace \left(\frac{1}{6}\bar{m}_{ns}+\frac{1}{2}s\mu_N\kappa_n B\right) E^n_F k^n_{F,s}-\frac{1}{6}\Big(E^n_F\Big)^3 \left(\arcsin\left(\frac{\bar{m}_{ns}}{E^n_F}\right)-\frac{\pi}{2}\right)\right. \cr
&-&\left. \bar{m}_{ns}^2\left(\frac{1}{2}s\mu_N\kappa_n B+\frac{1}{3}\bar{m}_{ns}\right)\ln\left|\frac{E^n_F+k^n_{F,s}}{\bar{m}_{ns}}\right|\right\rbrace. 
\eea

The differential magnetic susceptibility of nuclear matter is calculated from
the derivative of the magnetization with respect to the field $B$ for proton
and neutron, at constant baryonic density
\beq
\chi_n=\left.\frac{\partial {\cal M}_{n}}{\partial B}\right|_{\rho}\, .
\eeq
For the proton we obtain the  expression
\bea
\chi_p &=&\frac{q_p}{2\pi^2}\sum_{\nu,s}\left\{\frac{B E^p_F}{k^p_{F,\nu,s}}\left(\frac{q_p\nu}{\tilde m_{p \nu}}-s\mu_N\kappa_p \right)^2-2\bar{m}_{p\nu s} \left(\frac{q_p\nu}{\tilde m_{p \nu}}-s\mu_N\kappa_p \right)\ln\left|\frac{ E^p_F+k^p_{F,\nu,s}}{\bar m_{p\nu s}}\right|\right.\cr
&-&\left. B\left[\left(\frac{q_p\nu}{\tilde m_{p \nu}}-s\mu_N\kappa_p
    \right)^2-\bar{m}_{p\nu s}\frac{(q_p\nu)^2}{\tilde m_{p \nu}^3}
  \right] \ln\left|\frac{E^p_F+k^p_{F,\nu,s}}{\bar m_{p\nu
        s}}\right|\right\rbrace 
\label{chip}
\eea
whereas for the neutron we have
\bea
\chi_n
=\frac{1}{4\pi^2}\sum_{s}\left(s\mu_N\kappa_n\right)^2\left\lbrace
  E^n_F k^n_{F,s}+\bar{m}_{ns}(\bar{m}_{ns}+2 s \mu_N\kappa_n B)
  \ln\left|\frac{E^n_F+k^n_{F,s}}{\bar{m}_{ns}}\right|\right\rbrace. 
\label{chin}
\eea
At small values of the magnetic field B,  we derive for the magnetization expressions similar
to the ones derived in the non relativistic approach of Ref.\ \cite{aguirre2011}. We get for the proton spectrum  
\bea
E^p_{\nu s}&\simeq& \frac{{k^2_{z\nu s}}}{2 m^*}+ m^*+\mu_N B\left[2\frac{m}{m^*}n+\frac{m}{m^*}-s\left(\kappa_p+\frac{m}{m^*}sgn(q) \right)\right] \ ,
\eea
where $k_{z\nu s}$ is the component of the momentum parallel to the magnetic field. For the proton energy density we obtain in this limit
\bea
\varepsilon_p\simeq \frac{q_p B}{2\pi^2}\sum_{\nu,s}\left\lbrace\frac{{k^p_{F,\nu, s}}^3}{6 m^*}+ m^* k^p_{F,\nu,s}+\mu_N B\left[2\frac{m}{m^*}\nu + \frac{m}{m^*}-s\left(\kappa_p + \frac{m}{m^*} sgn(q_p)\right) \right] k^p_{F,\nu, s}\right\rbrace.  
\eea
Therefore, the expression the proton magnetization reads
\bea
{\cal M}_p &=&-\left.\frac{\partial \varepsilon_p}{\partial B}\right|_{\rho, Y_p} \cr
    &\simeq& -\frac{\varepsilon_p}{B} - \frac{q_p B}{2\pi^2}\sum_{\nu,s}\left\lbrace \mu_N \left[2\frac{m}{m^*}\nu + \frac{m}{m^*}-s\left(\kappa_p + \frac{m}{m^*} sgn(q_p)\right) \right] k^p_{F,\nu, s}\right\rbrace \cr
&\simeq&- 2 \frac{q_p B}{2\pi^2}\sum_{\nu,s}\left\lbrace \mu_N \left[2\frac{m}{m^*}\nu + \frac{m}{m^*}-s\left(\kappa_p + \frac{m}{m^*} sgn(q_p)\right) \right] k^p_{F,\nu, s}\right\rbrace\cr
&& -\frac{q_p }{2\pi^2}\sum_{\nu,s}\left\lbrace\frac{{k^p_{F,\nu, s}}^3}{6 m^*}+ m^* k^p_{F,\nu, s}\right\rbrace,
\eea
from which, finally, we obtain the following approximated expression  
\beq
{\cal M}_p = 2 \mu_N \left[\bar \kappa_p \hbox{W}_p -2 L-n_p\right], 
\eeq
where the quantities $\bar \kappa_p, \hbox{W}_p, L$ and $n_p$ are defined as
\bea
\bar \kappa_p&=&\kappa_p+\frac{m}{m^*}\ , \,\,\, \hbox{W}_p=\frac{q_p B}{2\pi^2}\sum_{\nu,s} s k^p_{F,\nu, s} \\
L&=&\frac{q_p B}{2\pi^2}\sum_{\nu,s} \frac{m}{m^*}n k^p_{F,\nu,s} \ , \,\,\, n_p=\frac{q_p B}{2\pi^2}\sum_{\nu,s} \frac{m}{m^*} k^p_{F,\nu,s} \ .
\eea
For the neutron magnetization we proceed in a similar way. Some details are given in the following. The interested reader is referred to Ref.\ \cite{APG11} 
for a complete derivation. In the same fashion at small B, we get for the neutron spectrum  
\bea
E^n_{s}&\simeq& m^*+\frac{k^2_s}{2 m^*}-s\mu_N\kappa_n B =m^*+\frac{k^2_{z,s}+k^2_{\perp, s}}{2 m^*} -s\mu_N \kappa_n B \ ,
\eea
where $k_{z,s}$ and $k_{\perp, s}$ are the components parallel and orthogonal to the magnetic field. For the neutron energy density we obtain
\bea
\varepsilon_n\simeq \left\lbrace m^* \rho_n -\mu_N\kappa_n B W_n +\frac{1}{(2\pi)^2}\sum_{s}\int \frac{k^4_{s}}{m^*}dk_{s}\right\rbrace,  
\eea
with 
\beq
\displaystyle W_n=\frac{1}{2\pi^2}\sum_{s}\int s k^2_{s}dk_{s}
\eeq
being the spin asymmetry density for the neutron. The following expression 
\beq
{\cal M}_n =-\left.\frac{\partial \varepsilon_n}{\partial B}\right|_{\rho, Y_p} 
    \simeq \mu_N\kappa_n W_n.
\eeq
is obtained for the neutron magnetization.
We notice that this approximated expression was used for
$\beta$-equilibrated stellar matter in Ref.~\cite{dzg2013}.  

\section{Results and discussion}
\label{section3}
In the following we present and discuss the results obtained for nuclear
matter under a strong magnetic field within the FSUGold parametrization of the
NLW model~\cite{fsugold}.  We extend our analysis to baryon densities up to three times saturation
 density, and magnetic field intensities within the range
 $10^{15}\,\hbox{G}\leqslant B\leqslant 10^{19}\, \hbox{G}$ in order to identify
 the sensitiveness of the strength of the magnetic field.  As referred
 in the introduction, we will consider in the following fields $B\le
10^{19}$ G.  Although,  there is no evidence that fields  as
  high as $10^{19}$ G exist in the crust,  we include these
  values in the figures for low density cases for the sake of
  completeness.

Spin-polarized isospin asymmetric nuclear matter can be seen as an infinite nuclear system composed by protons and neutrons. Each particle species, $i=n,p$, has two different fermionic components: particles with spin-up ($\upa$) and particles with spin-down ($\dwa$), having number densities $\rho_i^{\upa}$ and $\rho_i^{\dwa}$, respectively. The degree of spin polarization of the system can be studied through the relative polarization of the particle species $i$, defined by
\beq
\Delta_i=\frac{\rho_i^{\upa}-\rho_i^{\dwa}}{\rho_i^{\upa}+\rho_i^{\dwa}}.
\label{Delt}
\eeq   
Note that for small values of the magnetic field, the relation $\Delta_i=\rho_iW_i$ is fulfilled, with the quantities $W_p$ and $W_n$ defined previously. Note also that the value $\Delta_n=\Delta_p=0$ corresponds to non-polarized ({\it i.e.,} $\rho^\upa_n=\rho^\dwa_n$ and  $\rho^\upa_p=\rho^\dwa_p$) matter, whereas $\Delta_n=\pm 1$ ($\Delta_p=\pm 1$) means that neutrons (protons) are totally polarized, {\it i.e.,} all the neutron (proton) spins are along along the same direction.

\begin{figure}[hbtp]
\includegraphics[width=0.57\linewidth]{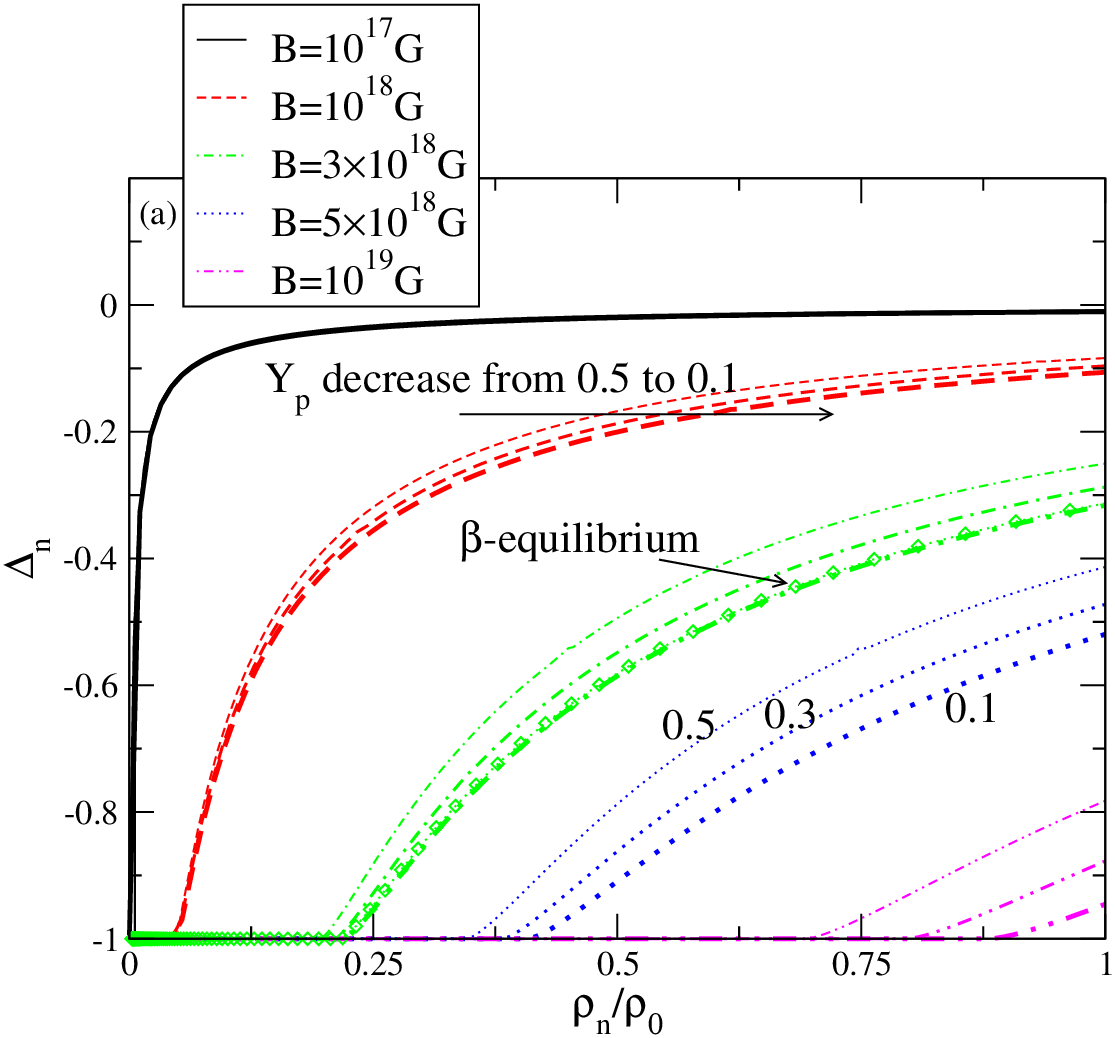}
\includegraphics[width=0.4\linewidth]{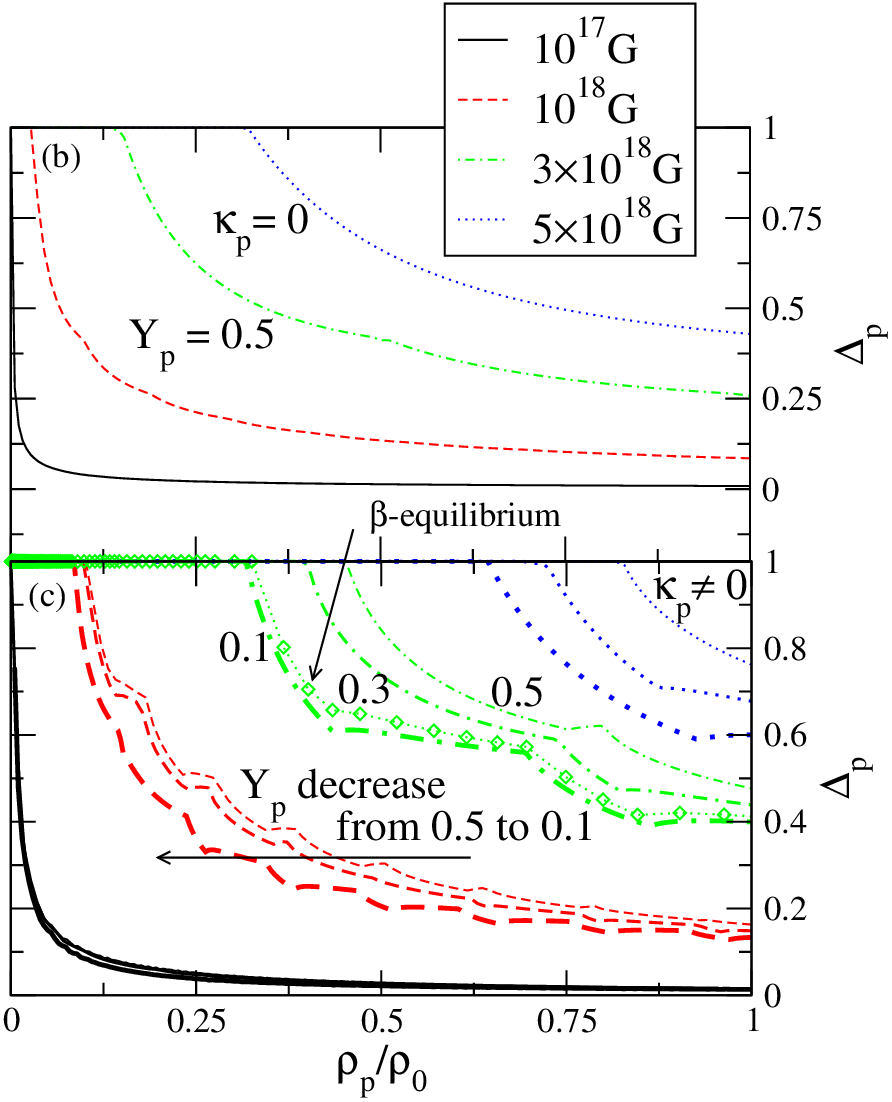}
\caption{(Color online) Neutron (left) and proton (right) relative polarization as a  function
  of the respective nucleon density, for several values of magnetic field
  and for $Y_p=0.1, 0.3, 0.5$ and for $\beta$-equilibrium matter. For protons it is shown the polarization not
  including (top) or including (bottom) the anomalous magnetic moment.}
\label{fig1}
\end{figure}

In Fig.~\ref{fig1} we show the neutron relative polarization $\Delta_{n}$ in terms of the neutron density (left panel), and the proton relative polarization $\Delta_{p}$ as function of the proton density (right panel). Results for both different magnetic field intensities from $B=10^{17}$ G to
$B=5\times 10^{18}$ G,  and different proton fraction, $Y_{p}=0.1,
0.3, 0.5$, are plotted. Decreasing proton fractions are depicted with
increasing line width. It should be pointed out that due to the fact that only 
the neutron or the proton densities are shown, changing the proton fraction 
changes also the total density: a given neutron density for neutron rich matter 
is attained at lower total densities than the same neutron density for symmetric matter.

We first discuss the results shown in  the left panel of Fig.~\ref{fig1} for the neutron polarization. In this range of fields, for very low densities, neutrons are totally polarized, (\textit{i.e.} $\Delta_n=-1$), up to a critical density above which they become partially polarized. This is in agreement with other calculations of pure neutron matter (see {\it i.e.,} Refs.\ \cite{APG08,aguirre2011, Aguirre13,abv2014}). Neutrons have always a negative polarization due to the different sign of its coupling to the electromagnetic field with respect to that of the protons. By increasing the value of the magnetic field at a fixed proton fraction the critical neutron density increases. It is also seen that the critical neutron density is larger for the more asymmetric and neutron richer matter, corresponding to a smaller
total nucleon density. The same is true for polarized proton matter, see  Fig.~\ref{fig1} right panel bottom: the total polarization occurs more easily in less dense matter, because the nucleon chemical potentials are smaller.

The critical neutron density depends on the proton fraction because changing $Y_p$ is equivalent to changing the neutron fraction ($Y_n=1-Y_p$), and, for a given neutron density, the larger the total baryonic density the more easily are neutrons totally polarized.

For a magnetic field of the order of $10^{17}$G neutrons are totally polarized only
at very small densities, and for neutron densities above $0.02\rho_0$ the partial
polarization of neutrons is below  10\%. This low degree of polarization is due to its weak anomalous magnetic moment.  

\begin{figure}[hbtp]
\includegraphics[scale=0.5]{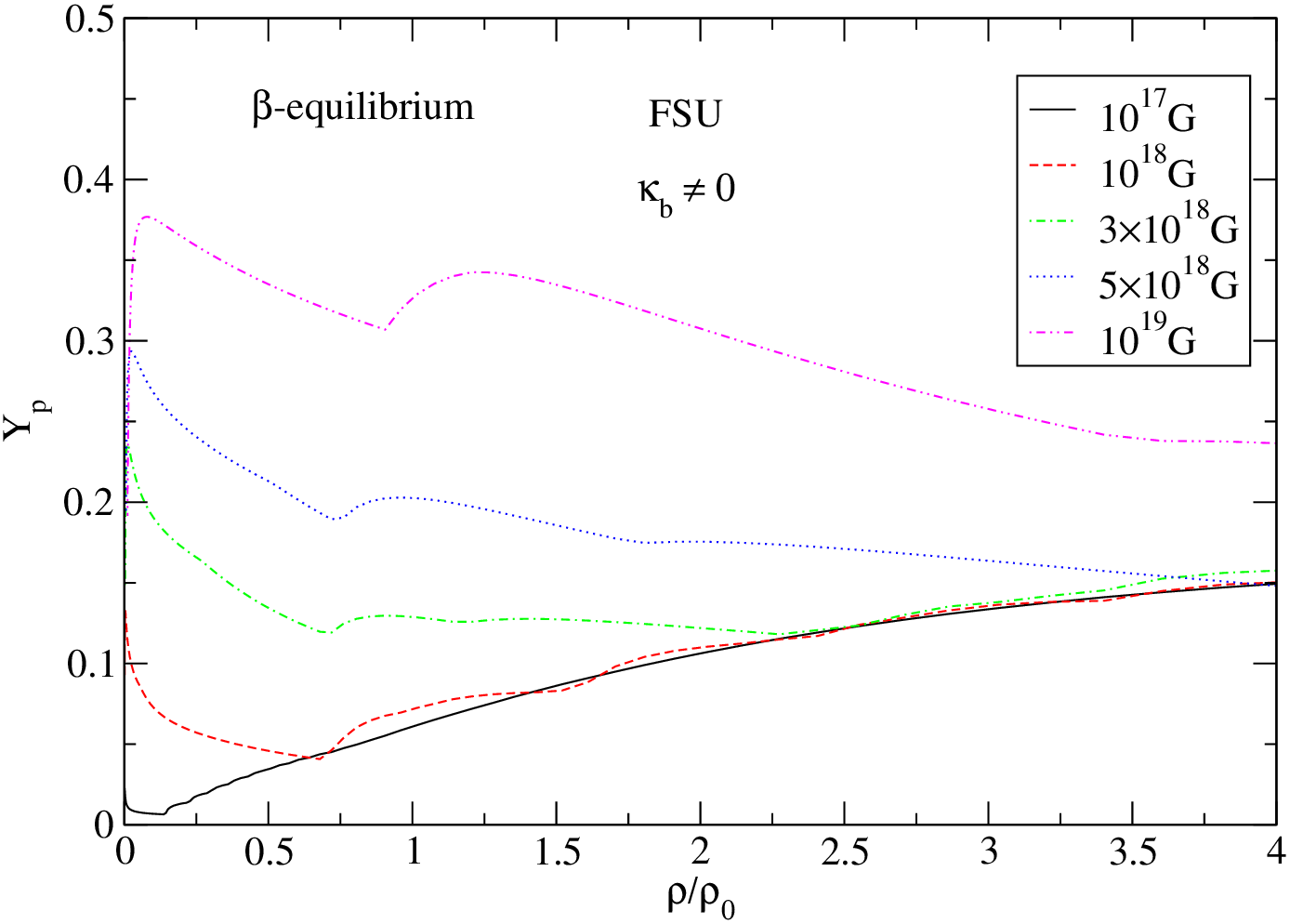}
\caption{(Color online) The proton fraction of $\beta$-equilibrium
  matter for several intensities of the magnetic field, taking into
  account the AMM. Results are obtained with FSUGold model.}
\label{fig10}
\end{figure}

\begin{figure}[hbtp]
\includegraphics[scale=.5]{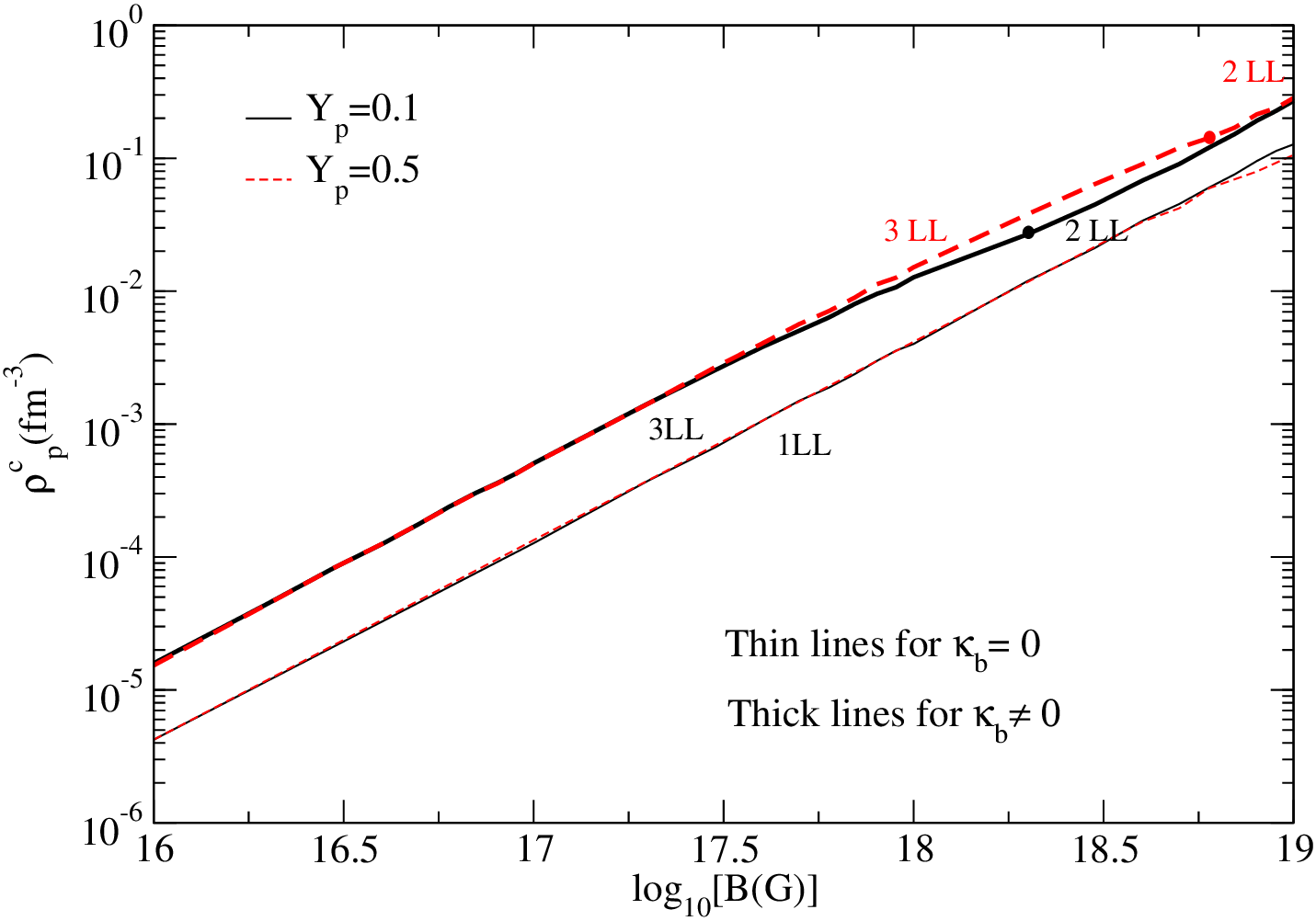}
\caption{(Color online) The critical density of protons as a function of the magnetic
  field, for $Y_{p}=0.1$ and $0.5$, and without/with AMM (thin/thick lines). The large
  dots indicate the filling of the indicated Landau level (LL)}
\label{fig2}
\end{figure}

In the panel (b) of Fig.~\ref{fig1} the relative proton polarization is
plotted for symmetric matter without AMM. Once more, for very low
densities, protons are totally polarized with $\Delta_p=1$, up to a critical density, 
where they become partially polarized with predominance of spin-up states, \textit{i.e.} $0<\Delta_p<1$. The
critical density increases with $B$. In the panel (c) the AMM is
included. The overall behavior does not change. Decreasing the  proton
fraction from $0.5$ to $0.1$, the critical proton density decreases,
associated with the increase of the total density. { We have included
in Fig.~\ref{fig1} a) and c), respectively,  the neutron and proton
polarizations for $\beta$-equilibrium matter in the presence of a
magnetic field $B=3\times 10^{18}$ G.  For neutrons the
$\beta$-equilibrium polarization is practically coincident with the
$Y_p=0.1$ results, and for protons, results for $Y_p=0.1$ are also
quite similar, in agreement with proton fraction expected for
$B=3\times 10^{18}$ G as shown in 
Fig.~\ref{fig10} where the proton fraction of $\beta$-equilibrium matter
for different magnetic field intensities is shown for $\rho<4 \rho_0$.}

In Fig.~\ref{fig2} we make a more careful analysis of the proton critical density as function
of the magnetic field, with and without AMM and for two values of the
proton fraction $Y_{p}=0.1$ and $0.5$. Protons are totally
  polarized on the region below the lines of critical proton
  density. If no AMM is included total polarization occurs when $\displaystyle eB >
\frac{{k^p_F}^2}{2}$, however, with AMM the following conditions should be satisfied,
$\displaystyle {k^p_F}^2<|4 \mu_N\kappa_n B|\sqrt{{m^*}^2+ 2eB}$ for stronger fields or
$\displaystyle {k^p_F}^2<|2 \mu_N\kappa_n B|\left( \sqrt{{m^*}^2+ 2eB}+\sqrt{{m^*}^2+
    4eB}\right)-2eB$ for weaker fields.
 The AMM favors the polarization so that the critical
density is  larger when the AMM is taken into account. For fields of
the order of $10^{16}$ G this difference is almost one order of
magnitude larger, while at $B=10^{19}$ G the difference reduces to a
factor of two. We conclude, therefore, that a realistic calculation
must include the AMM. The effect
of the isospin asymmetry on the critical proton density is related
with the filling of the Landau levels, which as referred before,
depends on the total baryonic density.
     
\begin{figure}[hbtp]
\includegraphics[scale=0.5]{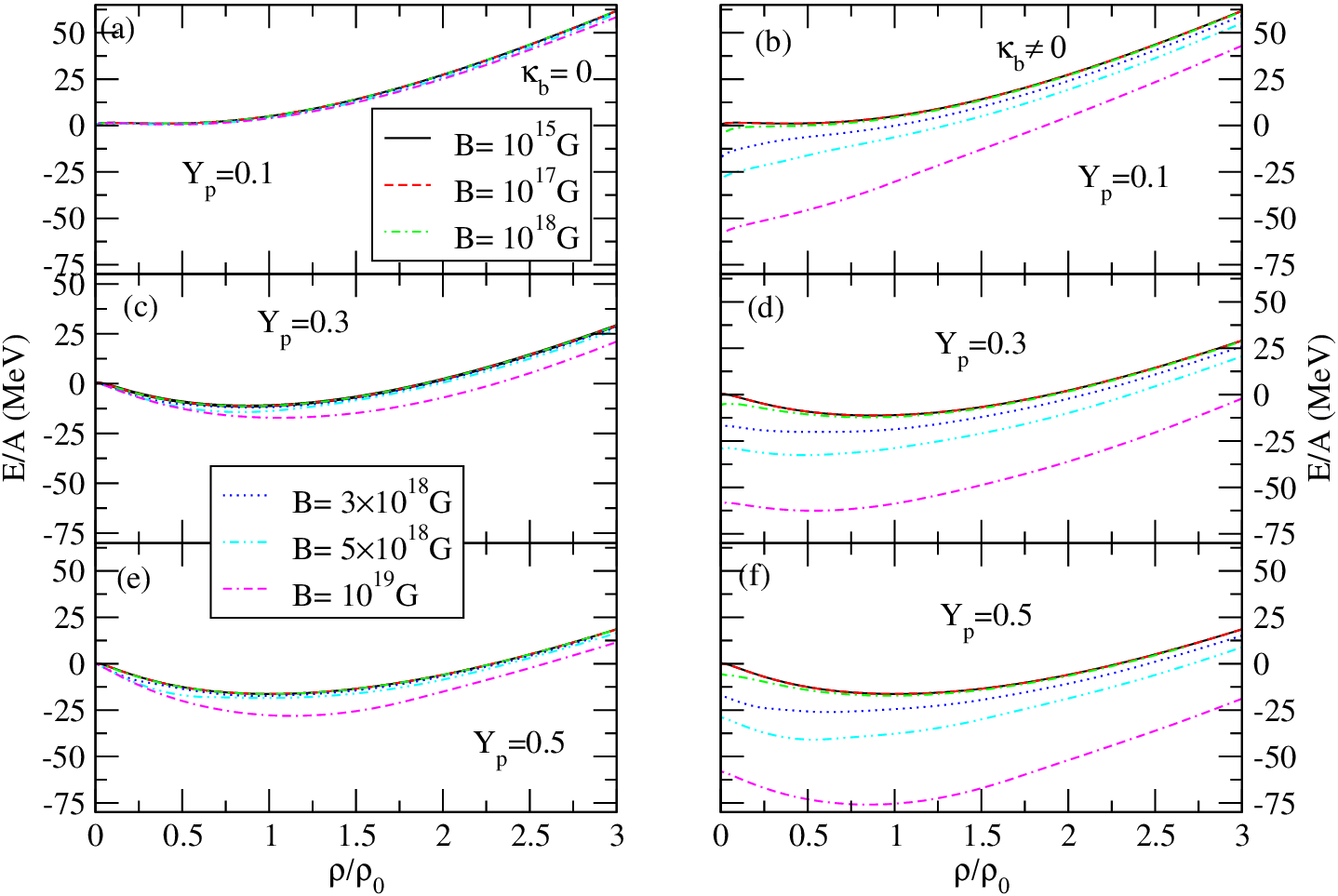}
\caption{(Color online) The energy per particle as a function of the density for
  several values of the magnetic field, and proton fractions. In the
  left panel no AMM was included. Results obtained with FSUGold model.}
\label{fig3}
\end{figure}

In Fig.~\ref{fig3} we show the energy per particle, defined by
$\displaystyle E/A=\frac{\varepsilon}{\rho}-m$, as a function of the
baryon density, for several values of the magnetic field, and for
$Y_{p}=0.1, 0.3$, and $0.5$. The curves obtained for $B< 10^{18}$ G are almost coincident.
In Ref.~\cite{aguirre2011}, results obtained with
$10^{14}<B<10^{18}$ G also practically coincide, however, they are
lower than the prediction for $B=0$. This could be due to some
difference in the parametrization or normalization in the
calculation done with $B=0$.

Ignoring the AMM (left panels), it is seen that  in average the effect of the magnetic field is to increase the binding of the nuclear matter, and the effect is stronger for more symmetric matter.  
The AMM has a strong effect for larger values of the magnetic field. Note that
for $B\geqslant 10^{18}$G there are already noticeable effects at densities $\rho\leqslant 0.25 \rho_0$.
For stronger magnetic fields, the effect of the AMM  leads clearly to an increase of the binding energy per particle.

\begin{figure}[hbtp]
\includegraphics[scale=.65]{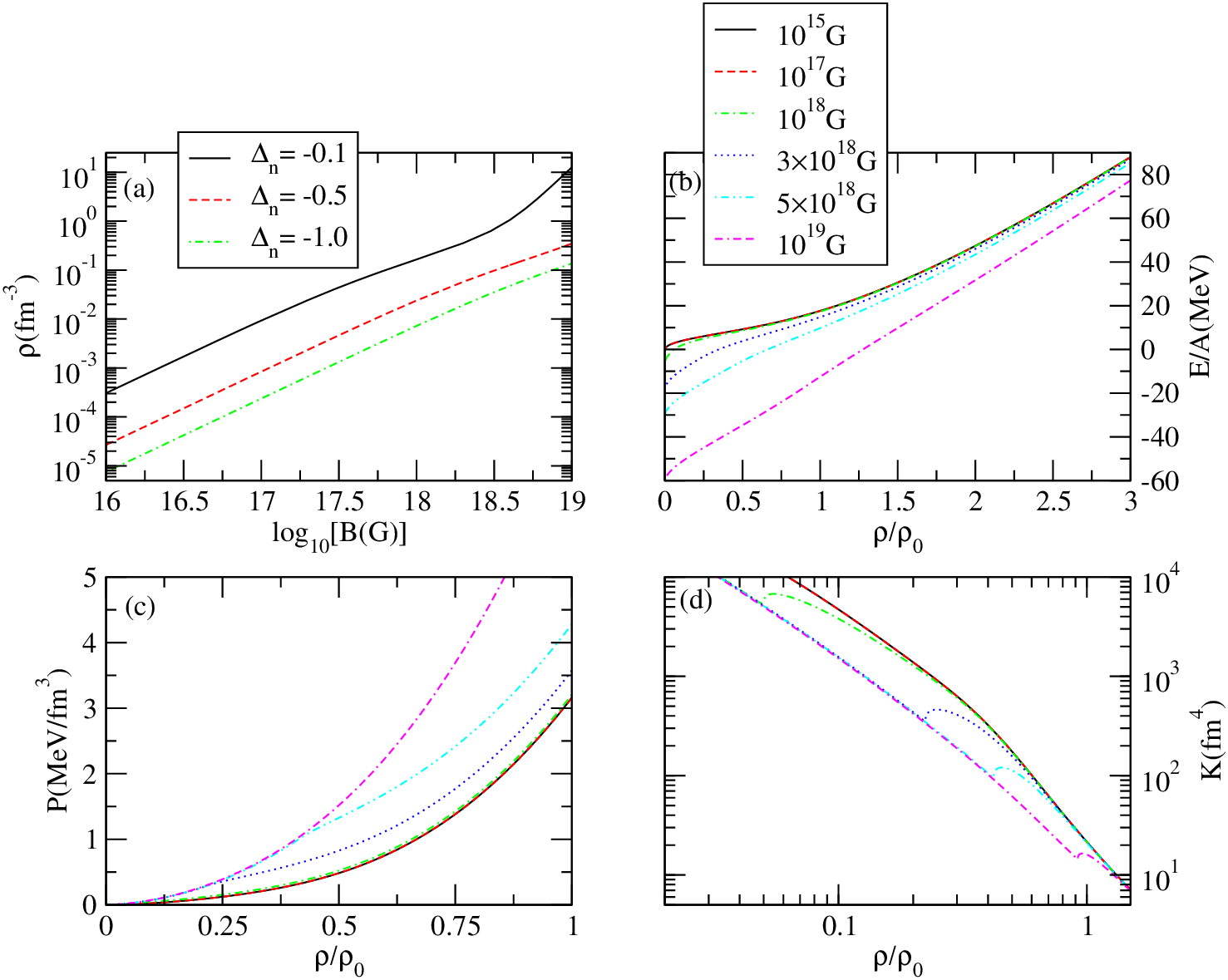}
\caption{(Color online) Several thermodynamic properties of neutron matter under the
  effect of an external magnetic field are shown: (a) the neutron critical
  density for total polarization and for partial polarization (50\%
  and 10\%) as a function of the
  magnetic field intensity, (b) the energy per particle, (c) the
  pressure, and (d) the compressibility $K$ are plotted as
  function of the total neutron density for several magnetic field
  intensities. Matter below the full, dashed and dot-dashed lines in
  panel a) have, respectively, $\Delta_n< -0.1, \, -0.5$ and
  $\Delta_n=-1$.}
\label{fig4}
\end{figure}

Several authors have studied neutron matter under the effect of strong magnetic fields within different frameworks and interactions. such as {\it e.g.,} the Gogny interaction \cite{ANG09a,ANG09b}. Recently, the authors of Ref.\ \cite{abv2014} have used both microscopic, namely the 
Brueckner-Hartree-Fock approach with the Argonne V18 nucleon-nucleon potential 
supplemented with a three body
force, and phenomenological approaches, in particular an effective
Skyrme model in a Hartree-Fock description and a
mean-field quantum hadrodynamical formulation with the FSUGold parametrization.  

In the following we will present neutron matter properties and discuss how they change
with the intensity of the magnetic field.
 In order to discuss the global state of polarization and bulk
 thermodynamical properties of pure neutron matter, we show in
 Fig.\ \ref{fig4}: (i) the  neutron critical density corresponding to
 the  transition from totally polarized as well as lines of partial
 polarization (50\% and 10\%) as a
 function of the magnetic field  (panel (a)), (ii) the energy per particle
 (panel (b)), (iii) the nucleonic pressure $P$ (panel (c)), and (iv) the
 compressibility $K$ (panel (d)), the last three quantities as a function of the total density for several magnetic field intensities.

 Just as before we have discussed for
 protons, the neutron critical density is an increasing function of
 the magnetic field and the critical density is defined imposing that
 the single particle energy of neutrons is smaller than the energy
 required to start populating the neutron spin up levels. This limit
 is defined by $\displaystyle \frac{{k^n_F}^2}{4m^*}<|\mu_N \kappa_n B|$ . A magnetic field $B=3 \times 10^{16}$ G has already a
 noticeable effect: a polarization of 10\% is expected at
 $\rho_n=0.001$ fm$^{-3}$ and 50\% for  $\rho_n=0.0001$ fm$^{-3}$. These are
 neutron densities of the order of the ones existing in the background
 neutron gas in the pasta phases of the inner crust.
 
\begin{figure}[hbtp]
\includegraphics[scale=0.65]{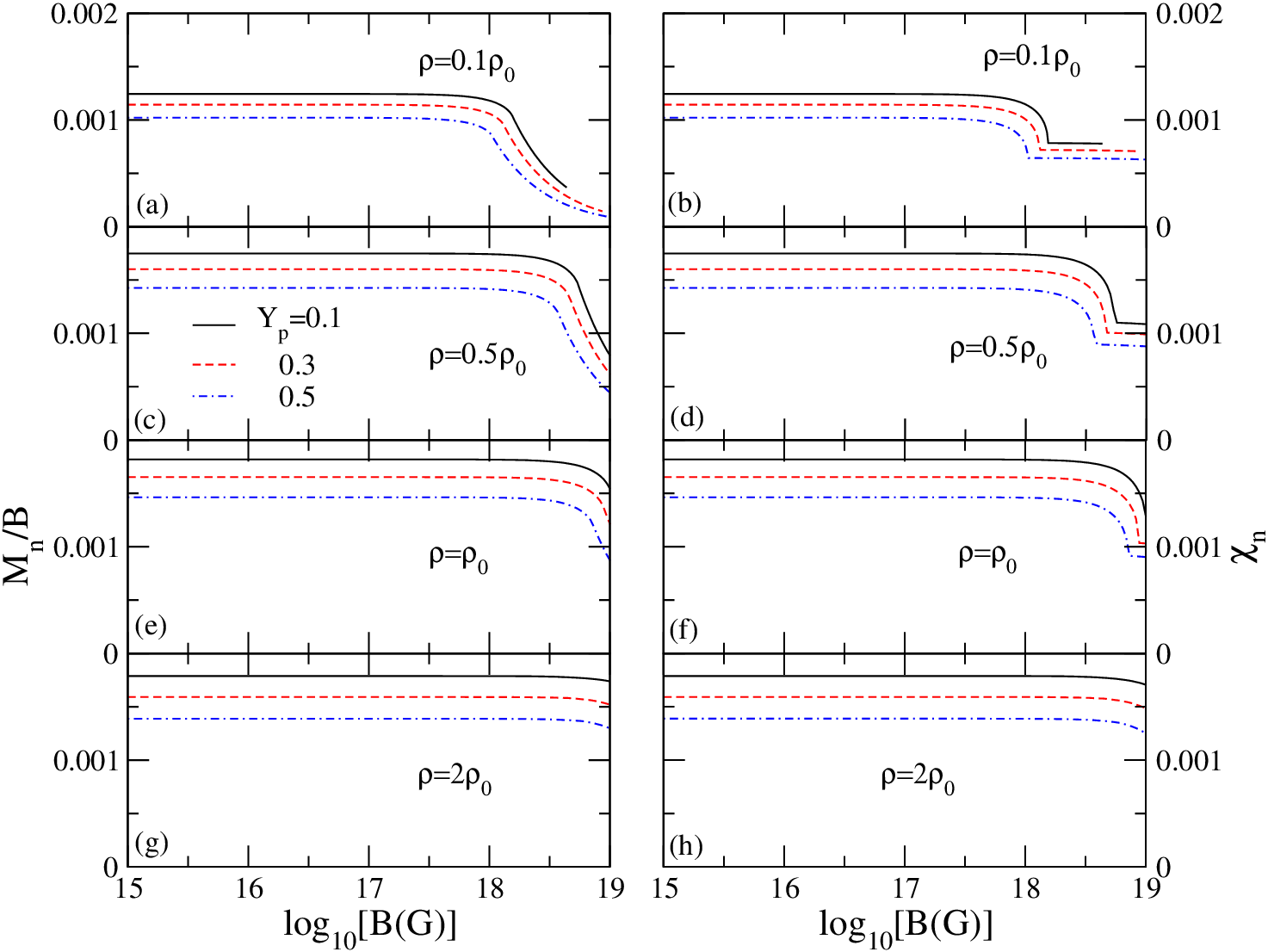}
\caption{(Color online) Neutron magnetic susceptibility (left panels) and differential
  susceptibility (right panels) as a function of the magnetic field and for
  several values of the density and isospin asymmetry. }
\label{fig5}
\end{figure}

From panel (b) of Fig.~\ref{fig4} , it is seen that the effects due to magnetic fields start
to be significant only for $B\geqslant 10^{18}$G at low baryon
densities. For $B\leqslant 10^{17}$G  neutron matter is not bound, as expected, because the magnetic field is too weak to have any effect on it. However, the
binding increases when the intensity of the magnetic field grows, and for $B=10^{19}$G pure neutron matter is
bound up to $\sim 1.5$ times saturation density. These results agree with the ones of
\cite{abv2014}.

The pressure in pure neutron matter is shown as a function of the
baryonic density  for several values of magnetic field in panel
(c). Only the region corresponding to densities below the saturation
density is plotted to show clearly the transition from totally
polarized to partially polarized neutron matter. The pressure
increases monotonically, showing, however, a softening at the
onset of partially polarized matter. This transition is clearly seen
on the isothermal compressibility $K$, defined through the first
derivative of the pressure, \textit{i.e.} $\displaystyle
1/K=\rho\frac{\partial P}{\partial \rho}$ (see panel (d)). For each
value of $B$, $K$ presents a kink at the critical density. 
 
\begin{figure}[hbtp]
\includegraphics[scale=0.5]{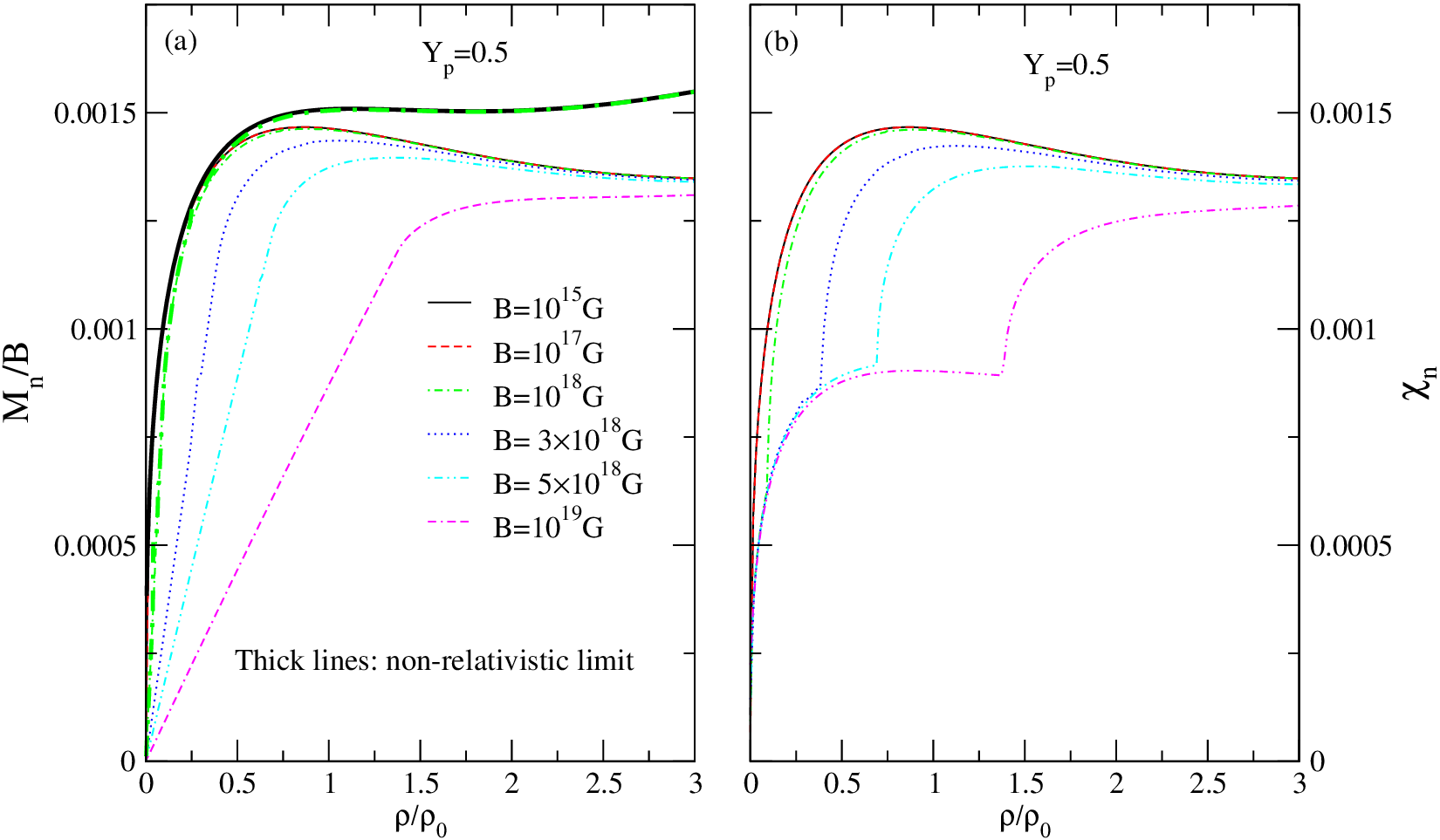}
\caption{(Color online) Neutron susceptibility (left) and differential susceptibility
  (right) as function of the baryon density, for several values of the
  magnetic field and for symmetric matter. The thick lines correspond
  to the non-relativistic limit for $B=10^{15}$ and $10^{18}$ G.}
\label{fig6}
\end{figure}

We next focus on the neutron magnetic susceptibility of asymmetric nuclear
matter. In the following we will consider both the magnetic
susceptibility defined by the ratio $M_{n}/B$ and the differential
susceptibility $\chi_{n}$. The dependence of the neutron magnetic
and differential susceptibility on the magnetic field intensity is shown on, respectively,
the left and right panels of Fig.~\ref{fig5}. Results are shown for different total
densities and proton fractions. As already found in Ref.~\cite{abv2014} the magnitude
 of the neutron susceptibility is very small, $\chi_n<0.0015$ for FSU.
Two different regimes are identified. In the low-field region corresponding to partially polarized
matter, $B\leqslant 3\times 10^{18}$G, $M_{n}/B$ and $\chi_{n}$ exhibit a
plateau. Beyond a threshold magnetic field, $M_{n}/B$ decreases,
showing a change of the slope, clearly seen as the kink
in $\chi_{n}$ which occurs at the  transition from totally polarized to partially polarized
matter. Above this critical magnetic field there is a strong decrease
of the susceptibility.

The neutron susceptibility decreases if the neutron fraction
decreases and has a non-monotonic behaviour with 
the density, increasing until $\sim \rho_0$ and
decreasing above this density for the lower magnetic field
intensities. This is clearly seen in Fig.~\ref{fig6} where we show 
the magnetic $M_{n}/B$ (panel (a)) and the differential $\chi_{n}$ (panel (b)) neutron susceptibility 
 as function of the baryon density,  for several values of the
magnetic field, and for a fixed proton fraction $Y_p=0.5$. At low densities $M_n/B$
practically does not change for $B<10 ^{18}$ G. For larger densities
this range increases to fields one order of  magnitude larger.
$M_{n}/B$  increases linearly with $\rho$ at low densities when
neutrons are totally polarized. At densities above the transition from totally polarized to 
partially polarized neutron matter, it continues increasing until a  plateau is
reached at high densities which corresponds to the limit when the
terms with the magnetic field $B$ are negligible,
$\displaystyle\chi_n\to\frac{1}{4 \pi^2} (\mu_N \kappa_n)^2 \left[E^n_F k^n_F + {m^*}^2\ln|\frac{E^n_F+k^n_F}{m^*}|\right].$
However, this value not always corresponds to
the maximum of the magnetization for weak fields. For
fields  below 10$^{18}$ G the maximum occurs at $\sim 0.5 - 1.0\rho_0$
before the plateau is attained.

\begin{figure}[hbtp]
\begin{tabular}{cc}
\includegraphics[width=0.41\linewidth]{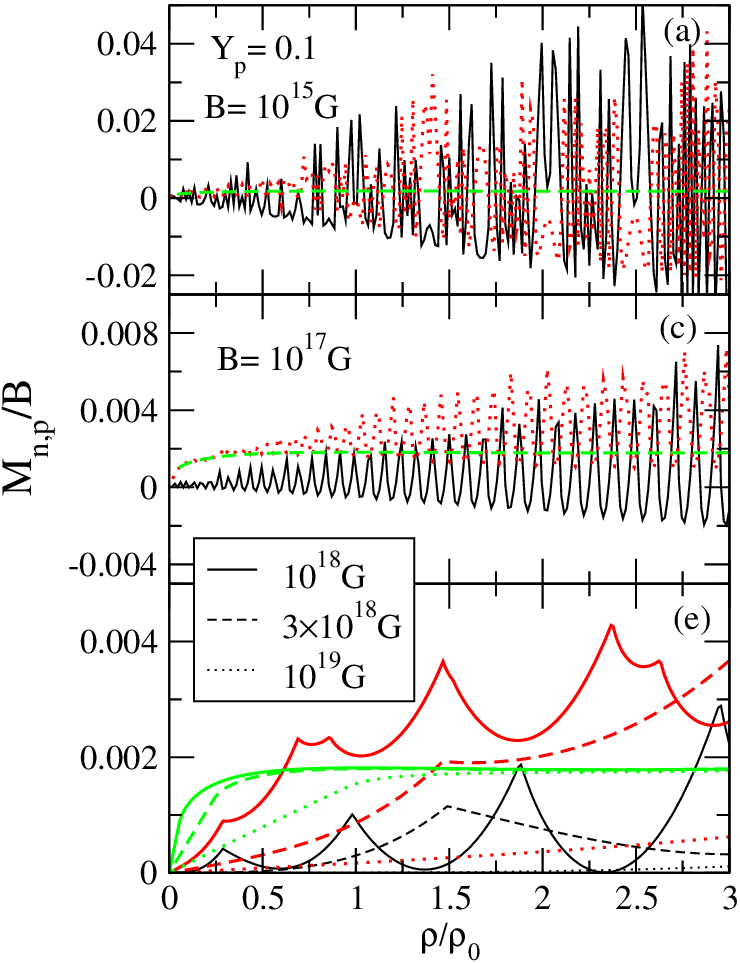}&
\includegraphics[width=0.4\linewidth]{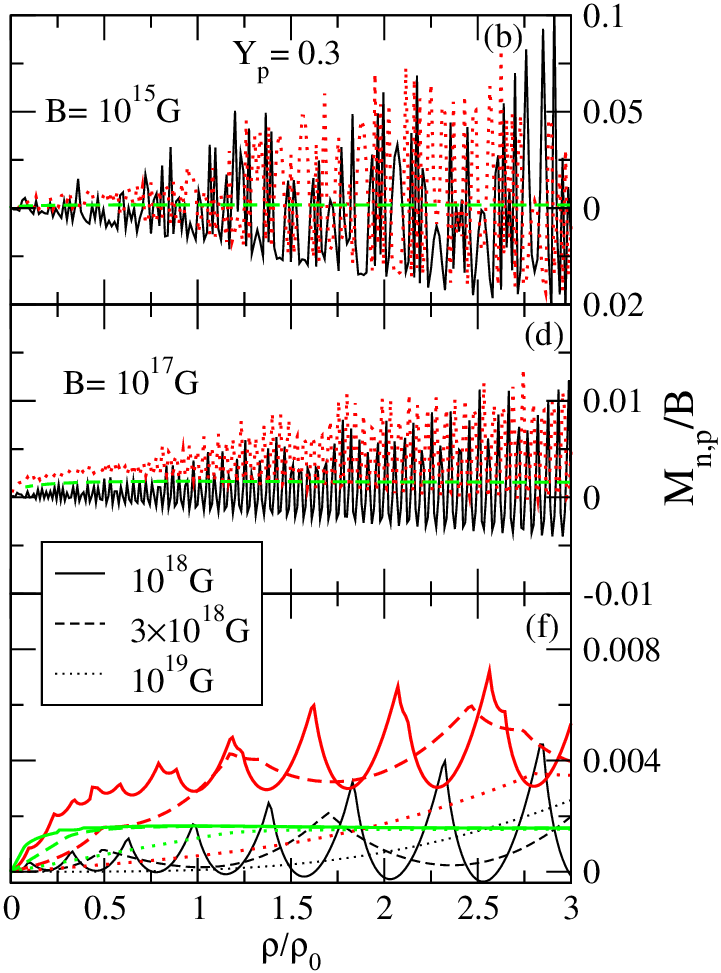}
\end{tabular}
\caption{(Color online) Proton and neutron  magnetic susceptibilities  as function of the density and
  for several values of the magnetic field. In the top and middle
  panels results with/without AMM are represented  by red dotted/black
  full lines for protons and green dashed lines for neutrons. In the
  bottom panel results with/without AMM for protons are given by the
  thick/thin lines and the green dashed lines are for neutrons.}
\label{fig7}
\end{figure}

The change of  slope seen in $M_{n}/B$ corresponds to the
kinks shown in the differential susceptibility $\chi_{n}$. We
also show the ratio $M_n/B$ in the non-relativistic limit, and contrary
to the relativistic result the susceptibility does not saturate but
increases monotonically with the density. The authors of Ref.~\cite{dzg2013}
have obtained this same increasing trend applying precisely the
non-relativistic limit of the  ratio $M_n/B$. 
A similar behaviour was also
obtained in Ref.~\cite{abv2014} with the Skyrme interaction, but in this
case it could be that this is due to the properties of most Skyrme
forces that  predict a phase transition to a ferromagnetic
phase at suprasaturation densities.

\begin{figure}[hbtp]
\begin{tabular}{cc}
\includegraphics[width=0.95\linewidth]{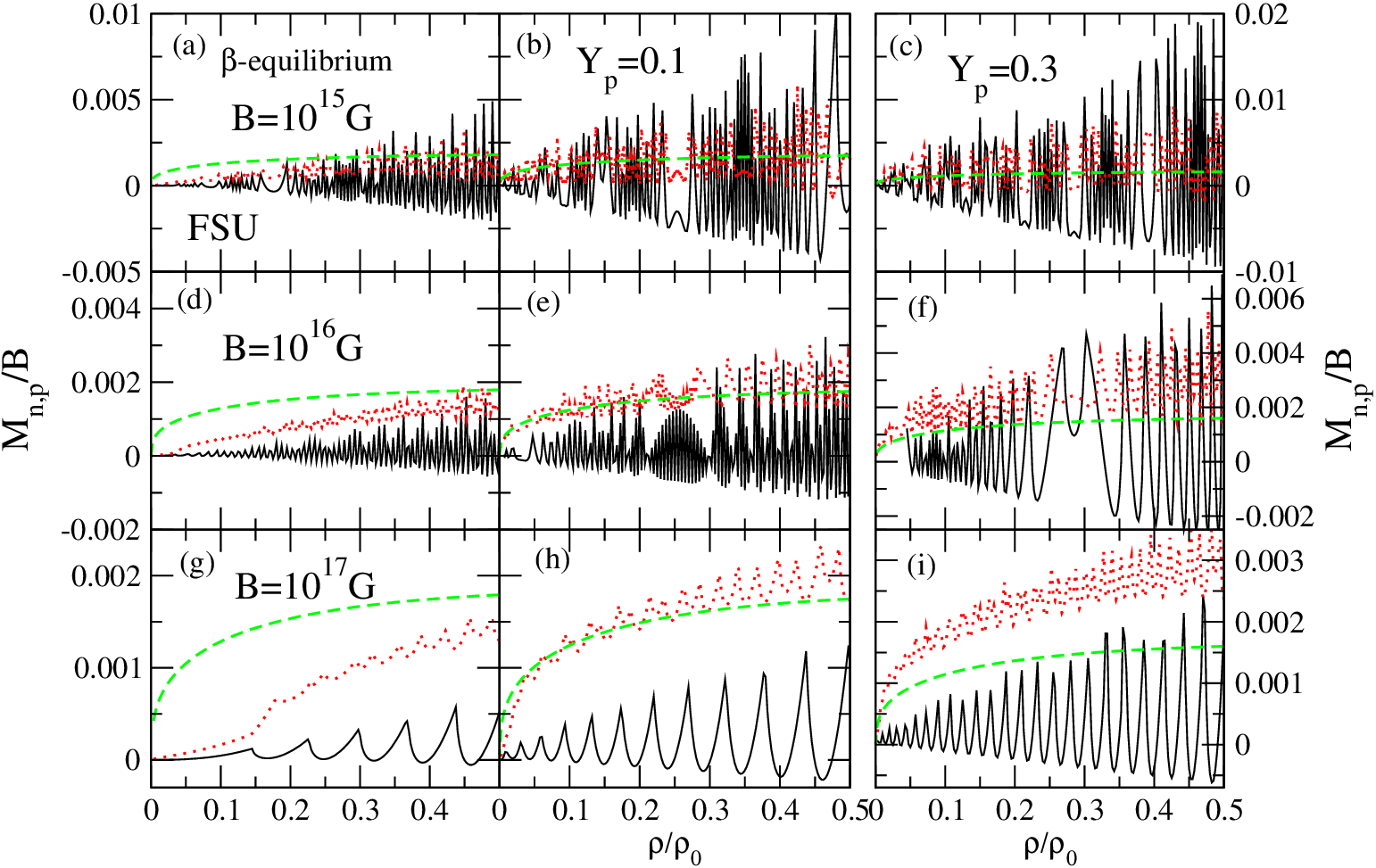} 
\end{tabular}
\caption{(Color online) Proton and neutron magnetic susceptibilities  as function of the density and
  for several values of the magnetic field, $B\le 10^{17}$ G and for
  densities of the interest for the neutron star crust. Results with/without AMM are represented  by red dotted/black
  full lines for protons and green dashed lines for neutrons.}
\label{fig7-1}
\end{figure}

We now discuss the proton magnetization and compare it to the neutron one.
The proton and neutron magnetic susceptibilities are presented in
Fig.~\ref{fig7} as a function of the  density for asymmetric nuclear
matter under different intensities of the magnetic fields and for two proton fractions,
$Y_p=0.1$ (left) and $0.3$ (right).
In the top and middle panels results for protons without  (black solid
  line) and with (red dotted line) AMM, and for neutrons (green lines) are
  shown respectively for $B=10^{15}$G and  $B=10^{17}$G.
In panels (e) and (f) the different curves are for $B=10^{18}$G (black solid
line), $3\times 10^{18}$G (red dashed line) and $10^{19}$G (blue
dotted line), without/with AMM (thin/thick lines). In all the panels
the neutron susceptibility is plotted with a green dashed curve. Please notice that the scale changes and the largest
susceptibilities occur for the smallest magnetic fields. 

\begin{figure}[hbtp]
\includegraphics[scale=0.6]{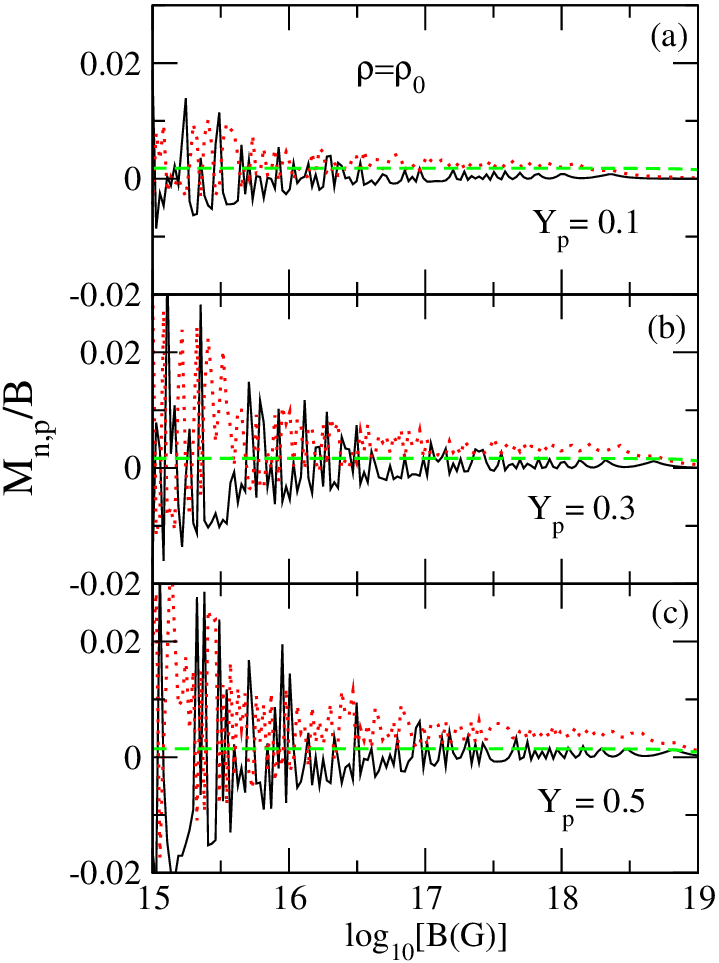}
\caption{(Color online) Proton and neutron magnetic susceptibilities and differential
  susceptibilities calculated at $\rho=\rho_0$ as function of the
  magnetic field for several values of the proton fraction, with/without AMM (red dotted/black
  full lines) for protons and for neutrons (dashed  green).}
\label{fig8}
\end{figure}

The susceptibility
curves for protons, at different magnetic fields, present the
well-known de Haas van Alphen oscillations associated to the change
in the number of Landau levels contributing at different fields. The filling of
the levels becomes more complicated if the AMM is included and this is
seen in the more complex structure of the  magnetic susceptibility
calculated with AMM, see bottom panel.
Some of the main conclusions that may be drawn from these figures are:
a) the proton magnetic susceptibility decreases much more strongly with B than the neutron magnetic susceptibility, and if $B$ changes from $10^{15}$ G to $10^{17}$ G 
its magnitude changes by almost an order of magnitude, and the same if
$B$ changes from $10^{17}$ G to $10^{19}$ G, while the
neutron susceptibility is practically unchanged; b) the inclusion of
the AMM may increase the proton susceptibility by a factor of two or more;
c) at low densities  the proton and neutron magnetic
susceptibilities are of the same order of magnitude and the fraction
of protons  defines how important is each contribution; d) for
$Y_p=0.1$ the neutron susceptibility is even larger than the proton
one for subsaturation densities, such as the ones occurring in the crust
of a neutron star.

In order to better understand the behaviour at low densities we plot 
in Fig.\ \ref{fig7-1} the magnetic susceptibilities for
subsaturation densities and fields $B \leq 10^{17}$G. These
are field intensities that could exist in the inner crust of a magnetar.  In fact,  in
the inner crust of a neutron star the background neutron gas will have densities that goes from zero to $\sim 0.5\rho_0$.
{  For neutron rich matter with $Y_p=0.1 $ and 0.3 or
 $\beta$-equilibrium matter  the neutron susceptibilities are of the
 same order of magnitude. There is, however, a clear difference
 between the proton susceptibilities: the larger the proton fraction
 the larger the susceptibility because the polarization is
 smaller. Total polarized matter has zero susceptibility. In
 particular, for these low fields $\beta$-equilibrium matter has a
 proton fraction below 0.1, see Fig.~\ref{fig10}, and, therefore, protons get more easily
 totally polarized. }

This is matter that is totally or partially polarized as can be seen
 looking at
Fig.~\ref{fig2} and \ref{fig4}:  pure neutron matter with
$\rho_n=0.0001$ fm$^{-3}$   is totally polarized by a field $B\sim
6\times 10^{16}$ G; for $\rho\sim 0.01$ 
fm$^{-3}$, $Y_p=0.1$ and $B=10^{17}$G the proton polarization is almost 100\%
but the neutron polarization is one order of magnitude smaller. The
magnetic susceptibility is larger for partially polarized matter and,
although, at $\sim 0.5 \rho_0$ the maximum neutron magnetic
susceptibility is attained for fields below $10^{18}$G, at $\sim
0.02 \rho_0$ it has already a magnitude that is half of the maximum value.
We expect that neutron polarization 
will affect the superfluidity of neutrons reducing its fraction. 

The effect of the proton fraction on the nucleon
susceptibilities is also seen 
in Fig.~\ref{fig8}  where these quantities are plotted as a function of the magnetic field intensity for several
proton fractions and for $\rho=\rho_0$. 
Neglecting the AMM contribution makes the proton
susceptibility go to zero as soon as only the first Landau level is
occupied, and protons are totally polarized. This is expected taking
the large $B$ limit of  Eq.~(\ref{chip}) with $\kappa_p=0$. For
$Y_p=0.1$ the neutron susceptibility is of the order of the proton one
or larger for $B>10^{17}$ G.
 Very strong oscillations occur for the weaker fields (or larger densities), and,
therefore the differential susceptibilities have not been shown, but if an averaging is done
as in~\cite{HHRS10} the average proton susceptibility would probably be of the
order of the neutron one also for these fields (and densities).
 We also conclude that it is important to
take into account AMM even for fields as small as 10$^{16}- 10^{17}$ G.

  We have obtained an overall  agreement  with the
conclusions obtained in previous works~\cite{bpl00,dzg2013}, in
particular, that the total magnetization decreases with an
increasing magnetic field and that its magnitude is quite small.

{  The effect of  the
magnetic field on the proton fraction of $\beta$-equilibrium matter
 was discussed in~\cite{APG11}:  
at $\rho=0.1\rho_0$  $Y_p$ increases from $Y_p< 0.005$ at
$10^{15}$G to 0.18 at $3.6\times 10^{18}$ G.  However, this effect is much
smaller at larger densities: for  $\rho=2\rho_0$, $4\rho_0$, $Y_p$  increases from $Y_p=0.15,\, 0.22$ at
$10^{15}$G to, respectively,   $Y_p=0.16$ and 0.23   at $3.6\times
10^{18}$ G. These results are in agreement with the proton fractions
plotted in Fig.~\ref{fig10}.
For the indicated  densities($0.1\rho_0, 2\rho_0, 4\rho_0$), the total
proton polarization is attained, respectively,  at Log$_{10}[B(G)]=16.2, \,
 18.4$ and 18.8. These results are confirmed in our present
work as particular cases of  a more systematic study,
see Fig. \ref{fig2}. Similar conclusions with respect to the effect of the magnetic field
on the proton fraction  have been drawn in \cite{rpp08},
where it is shown that, for $B=4.4 \times 10^{18}$G,
the magnetic field affects mainly densities below $2\rho_0$ and,
instead of the usual proton fractions below 0.1, proton fractions above this value are
expected. An increase of the proton fraction disfavors
 polarization, and, therefore, in $\beta$-equilibrium magnetized matter the
proton total polarization will not occur as easily.

In matter with neutrino trapping it was shown in \cite{rabhi2010} that at low
densities neutrino suppression occurs due to the larger proton fractions, which
can be as high as 0.4. Larger amounts of protons mean
that polarization  effects of the magnetic field on the protons will be smaller
and on neutrons larger. On the other hand the contribution of protons (neutrons)
to the total magnetization will increase (decrease).

At subsaturation densities matter is not
homogeneous and a pasta phase calculation is required. In \cite{lima2013} a first study was performed and it
was shown that fields below $10^{18}$G have a non negligible
effect on the pasta structure. In particular, it was shown that the
magnetic field disfavors neutron drip and, therefore, the neutron gas
outside the cluster is less dense for a given density and total
polarization occurs more easily.

In stellar $\beta$-equilibrium neutral matter besides protons and neutrons also the contribution of  electrons
(and muons above $~0.12$fm$^{-3}$) should be considered as discussed in
\cite{APG11,rpp08}. Consequently a complete description of stellar
matter requires also the leptonic contribution for the description of
quantities such as the  total magnetization, see \cite{dzg2013}.
}

\section{Summary and Conclusions}
\label{section4}

In the present work we have studied the proton and neutron polarization and magnetic
susceptibility of asymmetric nuclear matter within a relativistic
mean-field approach, in particular, the  FSUGold parametrization. 

The calculations were performed at a fixed proton fraction, and for
the proton results with and without AMM were compared. We have
calculated independently the proton and the neutron magnetic
susceptibilities and compared their magnitudes. Both of them are quite
small indicating that the magnetization induced by an external magnetic
field is weak. Similar conclusions have been obtained in \cite{bpl00,dzg2013}.

The proton
susceptibility 
oscillates very strongly due to the filling of Landau levels and decreases with an  increasing magnetic field. It was
shown that at subsaturations densities  the susceptibility calculated including the AMM may be
several times larger than the results obtained when it is ignored, for
magnetic fields with an intensity larger than $\sim 5\times 10^{16}$
G, and, therefore,  it is important to
take into account AMM for fields in the range 10$^{16}-10^{17}$ G.

The neutron susceptibility has a behaviour very  different not only
because it does not oscillate since the neutron has zero electric
charge but also because at large densities it converges to a value
that is independent of the magnetic field while the proton
susceptibility increases with the density for a fixed value of $B$. 
However, it was also shown that in the non-relativistic limit neutron
susceptibility increases monotonically with density.
We
have shown that at low density and for small proton fractions the
neutron susceptibility may be as large as the proton one  or even
larger.

We have also calculated the
transition density from  partially to  totally
polarized matter as a function of the magnetic field
intensity and it was shown that neutron matter is totally polarized by
a field $6\times 10^{16}$G and $\rho=0.0001$ fm$^{-3}$. 
The same field will also totally polarize the protons of asymmetric
nuclear matter at  $\rho=0.002$ fm$^{-3}$ with $Y_p=0.1$.
This behaviour
occurs for densities of relevance  in the neutron
star crusts and we expect that neutron superfluidity and transport properties of the crust will  be
affected by 
the presence of magnetic fields at least as strong as
$10^{16}$~G. This has been studied for the opacity {\it e.g.,} in
Ref.\ \cite{ANG10}.
In fact, at low densities, as the ones occurring in the inner crust it is
expected neutron superfluidity in the attractive channel $^1S_0$. A
partial or total neutron polarization will naturally hinder the
formation of neutron pairing. The consequences of the non existence or
reduction of neutron superfluidity would be a faster cooling of low
mass neutron stars, stars for which the direct Urca processes are not
expected, and a reduction of the glitch phenomena since the neutron
pairing determines the vortex structure
\cite{pairing,page2009,glitch}.  Also,  a reduction of the susceptibility would have strong effects on the mean free path
of a neutrino in dense matter, and, therefore, on the cooling of the
star \cite{fantoni2001}.


\section*{Acknowledgements}
One of the authors (A.~R.) want to acknowledge J. da Provid\^encia for many helpful and elucidating discussions. This work is partly supported by the project PEst-OE/FIS/UI0405/2014 developed under the inititative QREN financed by the UE/FEDER through the program
COMPETE-``Programa Operacional Factores de Competitividade'', and by ``NewCompstar'', COST Action MP1304.



\begin{thebibliography}{34}


\bibitem{epja} B. A. Li, A. Ramos, G. Verde and I. Vida\~na, Eds. {\it Topical issue on the nuclear symmetry energy}. Eur. Phys. J A {\bf 50}, issue 2 (2014).

\bibitem{baran} V. Baran ,  M. Colonna , V. Greco, and M. Di Toro, Phys. Rep. {\bf 410}, 335 (2005).

\bibitem{baoan} B. A. Li B, W. Chen, and C. M. Ko, Phys. Rep. {\bf 464}, 113 (2008).

\bibitem{steiner} A. W. Steiner, M. Prakash, J. Lattimer, and P. J. Ellis, Phys. Rep. {\bf 411}, 325 (2005).

\bibitem{harding06} 
A. K. Harding and D. Lai, Rep. Prog. Phys. 69, 2631 (2006). 

\bibitem{tatsumi06} T. Tatsumi, T. Maruyama, E. Nakano, and K. Nawa, Nucl. Phys. A {\bf 774}, 827 (2006).

\bibitem{thompson93} C. Thompson and Robert C. Duncan, Astrophys. J. {\bf 408}, 194 (1993).

\bibitem{BROWNELL} D. H. Brownell and J. Callaway, Nuovo Cimento B {\bf 60}, 169 (1969).

\bibitem{RICE} M. J. Rice, Phys. Lett. A {\bf 29}, 637 (1969).

\bibitem{CLARK} J. W. Clark and  N. C. Chao, Lett. Nuovo Cimento, {\bf 2}, 185 (1969).

\bibitem{CLARK2} J. W. Clark, Phys. Rev. Lett.  {\bf 23}, 1463 (1969).

\bibitem{SILVERSTEIN} S. D. Silvertein, Phys. Rev. Lett. {\bf 23}, 139 (1969).

\bibitem{OST} E. \O stgaard, Nucl. Phys. A {\bf 154}, 202 (1970).

\bibitem{PEAR} J. M. Pearson, G. Saunier, Phys. Rev. Lett. {\bf 24}, 325 (1970).

\bibitem{PANDA} V. R. Pandharipande, V. K. Garde, J. K. Srivastava, Phys. Lett. B {\bf 38}, 485 (1972).

\bibitem{BACK} S. O. B\"{a}ckman and C. G. K\"{a}llman, Phys, Lett. B {\bf 43}, 263 (1973).

\bibitem{HAENSEL} P. Haensel, Phys. Rev. C {\bf 11}, 1822 (1975).

\bibitem{JACK} A. D. Jackson, E. Krotscheck, D. E. Meltzer, and R. A. Smith, Nucl. Phys. A {\bf 386}, 125 (1982).

\bibitem{KUT} M. Kutschera and W. W\'ojcik, Phys. Lett. B {\bf 223}, 11 (1989); Phys. Lett. {\bf B 325} , 271 (1994).

\bibitem{MARCOS} S. Marcos, R. Niembro, M. L. Quelle, and J. Navarro, Phys. Lett. B {\bf 271}, 277 (1991);
                             P. Bernardos, S. Marcos, R. Niembro and M. L.. Quelle, Phys. Lett. B {\bf 356}, 175 (1996).

\bibitem{VIDAURRE} A. Vidaurre, J. Navarro, and J. Bernabeu, Astron. Astrophys. {\bf 135}, 361 (1984);
                                A. Rios, A. Polls and I. Vida\~na, Phys. Rev. C {\bf 71}, 055802 (2005).

\bibitem{RIOS} A. Rios, A. Polls and I. Vida\~na, Phys. Rev. C {\bf 71}, 055802 (2005).

\bibitem{GOGNY} D. L\'opez-Val, A. Rios, A. Polls, and I. Vida\~na, Phys. Rev. C {\bf 74}, 068801 (2006).

\bibitem{FANTONI} S. Fantoni, A. Sarsa, and K. E. Schmidt, Phys. Rev. Lett. {\bf 87}, 181101 (2001).

\bibitem{VIDANA} I. Vida\~na, A. Polls, and A. Ramos, Phys. Rev. C {\bf 65}, 035804 (2002).

\bibitem{VIDANAb} I. Vida\~na and I. Bombaci, Phys. Rev. C {\bf 66}, 045801 (2002).

\bibitem{VIDANAc} I. Bombaci, A. Polls, A. Ramos, A. Rios, and I. Vida\~na, Phys. Lett. B {\bf 632}, 638 (2006).


\bibitem{SAMMARRUCA1} F. Sammarruca and P. G. Krastev, Phys. Rev. C {\bf 75}, 034315 (2007).

\bibitem{SAMMARRUCA2} F. Sammarruca, Phys. Rev. C {\bf 83}, 064304 (2011).

\bibitem{BIGDELI} M. Bigdeli Phys. Rev. C {\bf 82} 054312 (2010).

\bibitem{ANG09a} M. A. P\'erez-Garc\'{i}a, J. Navarro, and A. Polls, Phys. Rev. C {\bf 80}, 025802 (2009).

\bibitem{ANG09b} M. A. P\'erez-Garc\'{i}a, Phys. Rev. C {\bf 80},
  045804 (2009).
\bibitem{cardall2001} C. Y. Cardall,  M. Prakash, and J. M. Lattimer, Astrophys. J. {\bf 554}, 322 (2001).
\bibitem{bpl02} A. Broderick, M. Prakash, and J. M. Lattimer,
  Phys. Lett. B 531, 167 (2002).
\bibitem{mcgill} S. A. Olausen, and V. M.  Kaspi,  Astrophysical
  Journal Supplement 212,  6 (2014).
 \bibitem{turolla2011} R. Turolla, S. Zane, J. A. Pons, P.  Esposito,  and N.  Rea,  ApJ  740, 105 (2011).
 \bibitem{rezzolla2012}J. Frieben and L. Rezzolla, MNRAS 427, 3406 (2012).
 \bibitem{oertel2014} Debarati Chatterjee, Thomas Elghozi, J\'er\^ome  Novak and  Micaela Oertel,  MNRAS  447, 3785 (2015).
 \bibitem{ruderman1998} M. Ruderman, T. Zhu, and K. Chin, ApJ. 482, 267 (1998).
 

\bibitem{kha2002} V. R. Khalilov, Phys. Rev. D 65, 056001 (2002).

\bibitem{APG08}  M. A. P\'erez-Garc\'{i}a, Phys. Rev. C {\bf 77}, 065806 (2008).

\bibitem{aguirre2011} R. Aguirre, Phys. Rev. C {\bf 83}, 055804 (2011).

\bibitem{APG11} M. A. P\'erez-Garc\'{i}a, C. Provid\^{e}ncia, and A. Rabhi, Phys. Rev. C {\bf 84}, 045803 (2011).

\bibitem{Aguirre13} R. Aguirre and E. Bauer, Phys. Lett. B {\bf 721}, 136 (2013).

\bibitem{abv2014} R. Aguirre, E. Bauer, and I. Vida\~na, Phys. Rev. C {\bf 89}, 035809 (2014).

\bibitem{bh82} R D Blandford and L Hernquist, J. Phys. C: Solid State Phys. {\bf 15}, 6233  (1982).

\bibitem{bpl00} A. Broderick, M. Prakash, and J. M. Lattimer, Astrophys. J. {\bf 537}, 351 (2000).

\bibitem{rabhi2010} A. Rabhi, C. Provid\^encia, J. Phys. G37, 075102 (2010) 
\bibitem{lima2013} R. C. R. de Lima, S. S. Avancini, and  C. Provid\^encia, Phys. Rev. C 88, 035804 (2013).

\bibitem{dzg2013} J. Dong, Wei Zuo and Jianzhong Gu, Phys. Rev. D {\bf 87}, 103010 (2013).

\bibitem{fsugold} B. G. Todd-Rutel and J. Piekarewicz,
  Phys. Rev. Lett. {\bf 95}, 122501 (2005).

\bibitem{fsu2} J. Piekarewicz, F. J. Fattoyev, and C. J. Horowitz,
Phys. Rev. C 90 015803 (2014); F. J. Fattoyev and J. Piekarewicz,
Phys. Rev. Lett. 111 162501 (2013); Matthias Hempel, Veronica Dexheimer, Stefan Schramm, and Igor Iosilevskiy,
Phys. Rev. C 88 014906 (2013);Neha Gupta and P. Arumugam,
Phys. Rev. C 88 015803 (2013); F. J. Fattoyev, J. Carvajal, W. G. Newton, and B. A. Li,
Phys. Rev. C 87 015806 (2013); S. S. Avancini, C. C. Barros, L. Brito, S. Chiacchiera, D. P. Menezes, and C. Provid\^encia,
Phys. Rev. C 85 035806 (2012); Fabrizio Grill, Constan軋 Provid\^encia, and Sidney S. Avancini,
Phys. Rev. C 85 055808 (2012).

\bibitem{cbp97} S. Chakrabarty, Phys. Rev. D 54, 1306 (1996); Somenath Chakrabarty, Debades Bandyopadhyay, and Subrata Pal,
Phys.Rev.Lett. 78, 2898-2901, (1997).

\bibitem{rpp08} A. Rabhi, C. Provid\^{e}ncia, and J. da
  Provid\^{e}ncia,  J. Phys. G: Nucl. Part. Phys. 35, 125201 (2008).
\bibitem{HHRS10} Xu-Guang Huang, Mei Huang, Dirk H. Rischke, and Armen Sedrakian, Phys. Rev. D 81, 045015 (2010).

\bibitem{ANG10} M. \'Angeles P\'erez-Garc\'{i}a, Eur. Phys. J. A {\bf 44}, 77 (2010).
\bibitem{pairing} M. Baldo, E. E.  Saperstein, and S. V.
  Tolokonnikov, Nucl. Phys. A, 749, 42 (2005); D. J. Dean, and
  M. Hjorth-Jensen, Rev. Mod. Phys., 75, 607–656, (2003).

\bibitem{page2009} D. Page, J. M. Lattimer, M. Prakash, and A. W. Steiner,
Astrophys. J. 707, 1131 (2009).
\bibitem{glitch}  B. Carter, in Lectures Notes in Physics, edited by
  D. Blaschke, N. K. Glendenning, and A. Sedrakian (Springer, Berlin,
  2000), Vol. 578, p. 54; A. Sedrakian, Phys. Rev. D 71, 083003
  (2005).
\bibitem{fantoni2001} S. Fantoni, A. Sarsa, and K. E. Schmidt, Phys. Rev. Lett. 87, 181101 (2001).

\end{thebibliography}
\end{document}